\documentclass[11pt,fleqn]{article}

\usepackage{a4}
\usepackage{amstext}
\usepackage{amsfonts}
\usepackage{amssymb}
\usepackage{color}
\usepackage{epsfig}
\usepackage{verbatim}
\usepackage[hang,small,bf]{caption}

\newcommand{\gtapprox}{\raisebox{-0.5ex}{$\,\stackrel{>}{\scriptstyle\sim}\,$}}
\newcommand{\ltapprox}{\raisebox{-0.5ex}{$\,\stackrel{<}{\scriptstyle\sim}\,$}}

\begin{document}

% ********************
% ********************
% ********************
% ********************
% ********************

\begin{center}

{\huge \bf The static-light baryon spectrum from}

{\huge \bf twisted mass lattice QCD}

\vspace{0.5cm}

SFB/CPP-11-20, HU-EP-11/20

\vspace{0.5cm}

\textbf{Marc Wagner, Christian Wiese} \\
Humboldt-Universit\"at zu Berlin, Institut f\"ur Physik, Newtonstra{\ss}e 15, D-12489 Berlin, Germany  \\

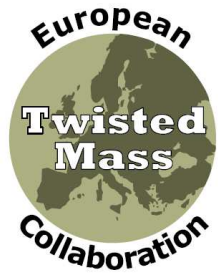
\vspace{0.7cm}

\begin{picture}(0,0)%
\includegraphics{Logo.pstex}%
\end{picture}%
\setlength{\unitlength}{4144sp}%
\begingroup\makeatletter\ifx\SetFigFont\undefined%
\gdef\SetFigFont#1#2#3#4#5{%
  \reset@font\fontsize{#1}{#2pt}%
  \fontfamily{#3}\fontseries{#4}\fontshape{#5}%
  \selectfont}%
\fi\endgroup%
\begin{picture}(1620,1620)(1,-781)
\end{picture}%

\vspace{0.4cm}

April~26, 2011

\end{center}

\vspace{0.1cm}

\begin{tabular*}{16cm}{l@{\extracolsep{\fill}}r} \hline \end{tabular*}

\vspace{-0.4cm}
\begin{center} \textbf{Abstract} \end{center}
\vspace{-0.4cm}

We compute the static-light baryon spectrum by means of Wilson twisted mass lattice QCD using $N_f = 2$ flavors of sea quarks. As light $u/d$ valence quarks we consider quarks, which have the same mass as the sea quarks with corresponding pion masses in the range $340 \, \textrm{MeV} \ltapprox m_\mathrm{PS} \ltapprox 525 \, \textrm{MeV}$, as well as partially quenched $s$ quarks, which have a mass around the physical value. We consider all possible combinations of two light valence quarks, i.e.\ $\Lambda$, $\Sigma$, $\Xi$ and $\Omega$ baryons corresponding to isospin $I \in \{ 0 \, , \, 1/2 \, , \, 1 \}$ and strangeness $S \in \{ 0 \, , \, -1 \, , \, -2 \}$ as well as angular momentum of the light degrees of freedom $j \in \{ 0 \, , \, 1 \}$ and parity $\mathcal{P} \in \{ + \, , \, - \}$. We extrapolate in the light $u/d$ and in the heavy $b$ quark mass to the physical point and compare with available experimental results. Besides experimentally known positive parity states we are also able to predict a number of negative parity states, which have neither been measured in experiments nor previously been computed by lattice methods.

\begin{tabular*}{16cm}{l@{\extracolsep{\fill}}r} \hline \end{tabular*}

\thispagestyle{empty}

% ********************
% ********************
% ********************
% ********************
% ********************

\newpage

\setcounter{page}{1}

\section{Introduction}

In this work we report on a lattice computation of the spectrum of $b$ baryons made from a heavy $b$ quark and two light quarks, which are $u$, $d$ and/or $s$.

Experimentally five $b$ baryon states have been observed. While $\Lambda_b$ has first been detected quite some time ago, $\Sigma_b$, $\Sigma_b^\ast$, $\Xi_b$ and $\Omega_b$ have only been discovered recently \cite{:2007rw,:2007ub,:2007un,Abazov:2008qm,Aaltonen:2009ny}. For the mass of $\Omega_b$ there are two different results, which are not in agreement.

On the theoretical side there are a number of lattice studies of the spectrum of $b$ baryons. Some of these consider static heavy quarks \cite{Michael:1998sg,Detmold:2007wk,Burch:2008qx,Detmold:2008ww,Lin:2009rx,Lin:2010wb} using Heavy Quark Effective Theory (HQET) (cf.\ e.g.\ \cite{Neubert:1993mb,Mannel:1997ky}), while others apply heavy quarks of finite mass \cite{Lewis:2008fu,Na:2008hz,Meinel:2009vv} mainly by means of Non-Relativistic QCD (cf.\ e.g.\ \cite{Thacker:1990bm}). For a recent review of lattice results on $b$ baryon masses cf.\ \cite{Lewis:2010xj}.

In this work we treat the $b$ quark in leading order of HQET, which is the static limit. In this limit there are no interactions involving the spin of the heavy quark, i.e.\ states are doubly degenerate. Therefore, it is common to label static-light baryons by integer spin/angular momentum $j$ and parity $\mathcal{P}$ of the light degrees of freedom. For the two light quarks we consider all possible combinations of $u$, $d$ and $s$, i.e.\ further quantum numbers are strangeness $S$ and isospin $I$. We use $N_f = 2$ flavors of dynamical quarks and study various ensembles with corresponding pion masses down to $\approx 340 \, \textrm{MeV}$. Our lattice spacing $a \approx 0.079 \, \textrm{fm}$ is rather fine and we use the Wilson twisted mass formulation of lattice QCD at maximal twist, which guarantees automatically $\mathcal{O}(a)$ improved spectral results. We compute all five experimentally known $b$ baryon states. We also make predictions for $\Xi'_b$, which has not yet been observed, as well as for a number of negative parity static-light baryons, which have neither been measured experimentally nor been computed by lattice methods.

The next-to-leading order of HQET, which removes the degeneracy with respect to the heavy quark spin, is $\mathcal{O}(1/m_Q)$, where $m_Q$ is the mass of the heavy quark. This correction is expected to be relatively small for $b$ baryons, e.g.\ experimentally the mass difference between $\Sigma_b$ and $\Sigma_b^\ast$ is only around $21 \, \textrm{MeV}$. Lattice methods to evaluate such $1/m_Q$ contributions have been established and tested in quenched studies of $B$ mesons \cite{Bochicchio:1991cy,Guazzini:2007bu,Blossier:2010jk,Blossier:2010vz}. We intend to explore these contributions using lattice techniques subsequently. An alternative way to predict the spectrum of $b$ baryons is to interpolate between charmed baryons, where the experimental spectrum is rather well known, and the static limit obtained by lattice QCD assuming a dependence as $1/m_Q$. Thus the splittings among $b$ baryons should approximately be $m_c / m_b \approx 1/3$ of those among the corresponding $c$ baryons.

We try to determine the $b$ baryon spectrum as fully as possible, i.e.\ we consider all possible light flavor combinations corresponding to $S \in \{0 \, , \, -1 \, , \, -2 \}$ and $I \in \{ 0 \, , \, 1 \}$ as well as both parity $\mathcal{P} = +$ and $\mathcal{P} = -$. This will help the construction of phenomenological models (cf.\ e.g.\ \cite{Ebert:2007nw}), might contribute to resolve open experimental issues (e.g.\ the above mentioned mass discrepancy for $\Omega_b$) and also provide valuable input for future experiments.

This study is in many aspects similar to our recent computation of the static-light meson spectrum \cite{Jansen:2008si,:2010iv}. Preliminary results have already been reported in conference proceedings \cite{Wagner:2010hj}.

The paper is organized as follows. In section~\ref{SEC486} we briefly recapitulate our lattice setup, which is discussed in more detail in~\cite{Jansen:2008si}. In section~\ref{SEC229} we discuss static-light baryon trial states, corresponding correlation matrices and how we extract the static-light baryon spectrum from these matrices as well as our extrapolation procedure to the physical $u/d$ quark mass. In section~\ref{SEC941} we interpolate between our static-light lattice results and experimental results for $c$ baryons, to account for the finite mass of the $b$ quark. We conclude with a brief summary and an outlook in section~\ref{SEC571}.

% ********************
% ********************
% ********************
% ********************
% ********************

\newpage

\section{\label{SEC486}Lattice setup}

In this work we use the same setup as for our recent computation of the static-light meson spectrum. For a more detailed presentation we refer to \cite{Jansen:2008si,:2010iv}.

We use $N_f = 2$ flavor gauge field configurations generated by the European Twisted Mass Collaboration (ETMC). The gauge action is tree-level Symanzik improved \cite{Weisz:1982zw},
\begin{eqnarray}
S_\mathrm{G}[U] \ \ = \ \ \frac{\beta}{6} \bigg(b_0 \sum_{x,\mu\neq\nu} \textrm{Tr}\Big(1 - P^{1 \times 1}(x;\mu,\nu)\Big) + b_1 \sum_{x,\mu\neq\nu} \textrm{Tr}\Big(1 - P^{1 \times 2}(x;\mu,\nu)\Big)\bigg)
\end{eqnarray}
with $b_0 = 1 - 8 b_1$ and $b_1 = -1/12$. The fermionic action is Wilson twisted mass (cf.\ \cite{Frezzotti:2000nk,Frezzotti:2003ni,Frezzotti:2004wz,Shindler:2007vp}),
\begin{eqnarray}
\label{EQN963} S_\mathrm{F}[\chi,\bar{\chi},U] \ \ = \ a^4 \sum_x \bar{\chi}(x) \Big(D_{\rm W} + i \mu_\mathrm{q} \gamma_5 \tau_3\Big) \chi(x) , 
\end{eqnarray}
where
\begin{eqnarray}
D_\mathrm{W} \ \ = \ \ \frac{1}{2} \Big(\gamma_\mu \Big(\nabla_\mu + \nabla^\ast_\mu\Big) - a \nabla^\ast_\mu \nabla_\mu\Big) + m_0 ,
\end{eqnarray}
$\nabla_\mu$ and $\nabla^\ast_\mu$ are the gauge covariant forward and backward derivatives, $m_0$ and $\mu_\mathrm{q}$ are the bare untwisted and twisted quark masses respectively, $\tau_3$ is the third Pauli matrix acting in flavor space and $\chi = (\chi^{(u)} , \chi^{(d)})$ represents the quark fields in the so-called twisted basis. The twist angle $\omega$ is given by $\tan(\omega) = \mu_\mathrm{R} / m_\mathrm{R}$, where $\mu_\mathrm{R}$ and $m_\mathrm{R}$ denote the renormalized twisted and untwisted quark masses. $\omega$ has been tuned to $\pi / 2$ by adjusting $m_0$ appropriately (cf.\ \cite{Boucaud:2008xu} for details). As argued in \cite{Jansen:2008si} this ensures automatic $\mathcal{O}(a)$ improvement for static-light spectral quantities, e.g.\ mass differences between static-light baryons and the lightest static-light meson (the ``$B$/$B^\ast$ meson''), the quantities we are focusing on in this work.

% a = 0.079(3)

% hbar c = 1 = 197.327 MeV fm

% > 0.1362 / 0.079 * 197.327
% [1] 340.2017
% > sqrt(0.0007^2 + (0.1362*0.003/0.079)^2) / 0.079 * 197.327
% [1] 13.03684
% a m_pi(mu = 0.0040) = 0.1362(7) --> 340(13) MeV

% > 0.1694 / 0.079 * 197.327
% [1] 423.129
% > sqrt(0.0004^2 + (0.1694*0.003/0.079)^2) / 0.079 * 197.327
% [1] 16.09922
% a m_pi(mu = 0.0064) = 0.1694(4) --> 423(16) MeV

% > 0.1940 / 0.079 * 197.327
% [1] 484.5752
% > sqrt(0.0005^2 + (0.1940*0.003/0.079)^2) / 0.079 * 197.327
% [1] 18.44392
% a m_pi(mu = 0.0085) = 0.1940(5) --> 485(18) MeV

% > 0.2100 / 0.079 * 197.327
% [1] 524.5401
% > sqrt(0.0005^2 + (0.2100*0.003/0.079)^2) / 0.079 * 197.327
% [1] 19.95836
% a m_pi(mu = 0.0100) = 0.2100(5) --> 525(20) MeV

% spatial lattice extension: L = 24 * 0.079(3) = 1.90(7) fm
% m_PS * L >= 0.1362(7) * 24 = 3.27(2)

The ensembles of gauge field configurations we are considering are listed in Table~\ref{TAB100}. They correspond to a single value of the lattice spacing $a \approx 0.079 \, \textrm{fm}$, but various values of the pion mass in the range $340 \, \textrm{MeV} \ltapprox m_\textrm{PS} \ltapprox 525 \, \textrm{MeV}$. The lattice extension is $L^3 \times T = 24^3 \times 48$, which amounts to $L \approx 1.9 \, \textrm{fm}$ and $m_\textrm{PS} L \gtapprox 3.3$. Details regarding the generation of these gauge field configurations and computation and analysis of standard quantities (e.g.\ lattice spacing or pion mass) can be found in \cite{Boucaud:2008xu,Baron:2009wt}.

\begin{table}[htb]
\begin{center}

\begin{tabular}{|c|c|c||c|c||c|}
\hline
 & & & & & \vspace{-0.40cm} \\
$\beta$ & $L^3 \times T$ & $\mu_\mathrm{q}$ & $a$ in $\textrm{fm}$ & $m_\textrm{PS}$ in $\textrm{MeV}$ & \# of gauges \\
 & & & & & \vspace{-0.40cm} \\
\hline
 & & & & & \vspace{-0.40cm} \\
$3.90$ & $24^3 \times 48$ & $0.0040$ & $0.079(3)$ & $340(13)$  & $200$           \\
       &                  & $0.0064$ &            & $423(16)$  & $\phantom{0}50$ \\
       &                  & $0.0085$ &            & $485(18)$  & $\phantom{0}50$ \\
       &                  & $0.0100$ &            & $525(20)$  & $\phantom{0}50$\vspace{-0.40cm} \\
 & & & & & \\
\hline
\end{tabular}
\caption{\label{TAB100}ensembles of gauge field configurations ($a$ and $m_\textrm{PS}$ have been taken from \cite{Baron:2009wt}; \# of gauges: number of gauge field configurations considered).}
\end{center}
\end{table}

We treat static-light baryons containing valence $s$ quarks in a partially quenched approach, where the mass of these quarks, $\mu_{\mathrm{q},\textrm{valence }s} = 0.022$, is approximately equal to the mass of the physical $s$ quark taken from a study of strange mesons using the same gauge field configurations \cite{Blossier:2007vv,Blossier:2009bx}. Note that partially quenched $s$ quarks can be realized in two ways, either with a twisted mass term $+i \mu_{\mathrm{q},\textrm{valence }s} \gamma_5$ or $-i \mu_{\mathrm{q},\textrm{valence }s} \gamma_5$ corresponding to the upper and the lower entry in the quark field doublet $\chi$ respectively. We consider both possibilities and denote them by $\chi = (\chi^{(s^+)} , \chi^{(s^-)})$.

In Table~\ref{TAB100} we also list the number of gauge configurations, on which we have computed static-light baryon correlation functions.

% ********************
% ********************
% ********************
% ********************
% ********************

\newpage

\section{\label{SEC229}The static-light baryon spectrum}

With static-light baryons we refer to baryons made from a single static quark and two light quarks, which can either be $u$, $d$ and/or $s$.

% **********

\subsection{Static-light baryon trial states}

% *****

\subsubsection{Static-light baryon creation operators in the continuum}

We start by discussing symmetries and quantum numbers of static-light baryons and corresponding creation operators in the continuum.

The continuum analogs of our lattice static-light baryon creation operators are
\begin{eqnarray}
\label{EQN001} \mathcal{O}_{\Gamma,\psi^{(1)} \psi^{(2)}}^\textrm{physical}(\mathbf{x}) \ \ = \ \ \epsilon^{a b c} Q^a(\mathbf{x}) \Big((\psi^{b,(1)}(\mathbf{x}))^T \mathcal{C} \Gamma \psi^{c,(2)}(\mathbf{x})\Big) ,
\end{eqnarray}
where $Q$ is a static quark operator and $\psi^{(n)}$ are light quark operators (in the usual physical basis). The upper indices $a$, $b$ and $c$ are color indices, $\mathcal{C} = \gamma_0 \gamma_2$ is the charge conjugation matrix and $\Gamma$ is a combination of $\gamma$ matrices, i.e.\ a $4 \times 4$ matrix acting in spin space.

Since there are no interactions involving the static quark spin, it is appropriate to label static-light baryons by the angular momentum of their light degrees of freedom $j$. For creation operators (\ref{EQN001}) it is determined by $\Gamma$ and can either be $j = 0$ or $j = 1$. $j = 0$ states correspond to total angular momentum $J = 1/2$, while $j = 1$ states correspond to degenerate pairs of states with total angular momentum $J = 1/2$ and $J = 3/2$, respectively.

% Parity is a matter of convention: it can be defined with +/- (even with an arbitrary phase); the definition, however, affects the resulting quantum number; moreover, baryons and antibaryons have opposite parity.

% According to Vladimir parity is usually defined according to
%   P \psi = -\gamma_0 \psi.
% Then for static quarks
%   Q --P--> +Q and \bar{Q} --P--> -\bar{Q}.
% Moreover, e.g.
%   proton --P--> +proton and antiproton --P--> -antiproton.

Parity is also a quantum number depending on $\Gamma$. Either $\mathcal{P} = +$ or $\mathcal{P} = -$.

The flavor quantum numbers are isospin $I$ and strangeness $S$. To access all possible combinations, we consider light quark flavors $\psi^{(1)} \psi^{(2)} = ud - du$ (corresponding to $I = 0$, $S = 0$), $\psi^{(1)} \psi^{(2)} \in \{ uu \, , \, dd \, , \, ud + du \}$ (corresponding to $I = 1$, $S = 0$), $\psi^{(1)} \psi^{(2)} \in \{ us \, , \, ds \}$ (corresponding to $I = 1/2$, $S = -1$) and $\psi^{(1)} \psi^{(2)} = ss$ (corresponding to $I = 0$, $S = -2$).

Creation operators $\mathcal{O}_{\Gamma,\psi^{(1)} \psi^{(2)}}^\textrm{physical}$ and the quantum numbers of their associated trial states $\mathcal{O}_{\Gamma,\psi^{(1)} \psi^{(2)}}^\textrm{physical} | \Omega \rangle$ are collected in Table~\ref{TAB002}. Note that certain $\Gamma,\psi^{(1)} \psi^{(2)}$ combinations do not need to be considered, since the corresponding creation operators are identical zero due to the anticommutation property of quark operators. Such $\Gamma,\psi^{(1)} \psi^{(2)}$ combinations are either omitted from the table or marked with ``X''.

\begin{table}[htb]
\begin{center}
\begin{tabular}{|c||c|c||c|c|c||c|c|c||c|c|c|}
\hline
 & & & & & & & & & & & \vspace{-0.40cm} \\
$\Gamma$ & $j^\mathcal{P}$ & $J$ & $I$ & $S$ & name & $I$ & $S$ & name & $I$ & $S$ & name \\
 & & & & & & & & & & & \vspace{-0.40cm} \\
\hline
 & & & & & & & & & & & \vspace{-0.40cm} \\
$\gamma_5$ & $0^+$ & $1/2$ & $0$ & $0$ & $\Lambda_b$ & $1/2$ & $-1$ & $\Xi_b$ & X & X & X \\
$\gamma_0 \gamma_5 $ & $0^+$ & $1/2$ & $0$ & $0$ & $\Lambda_b$ & $1/2$ & $-1$ & $\Xi_b$ & X & X & X \\
$1$ & $0^-$ & $1/2$ & $0$ & $0$ & & $1/2$ & $-1$ & & X & X & X \\
$\gamma_0$ & $0^-$ & $1/2$ & $1$ & $0$ & & $1/2$ & $-1$ & & $0$ & $-2$ & \\
 & & & & & & & & & & & \vspace{-0.40cm} \\
\hline
 & & & & & & & & & & & \vspace{-0.40cm} \\
$\gamma_j$ & $1^+$ & $1/2$, $3/2$ & $1$ & $0$ & $\Sigma_b$, $\Sigma_b^\ast$ & $1/2$ & $-1$ & & $0$ & $-2$ & $\Omega_b$ \\
$\gamma_0 \gamma_j$ & $1^+$ & $1/2$, $3/2$ & $1$ & $0$ & $\Sigma_b$, $\Sigma_b^\ast$ & $1/2$ & $-1$ & & $0$ & $-2$ & $\Omega_b$ \\
$\gamma_j \gamma_5$ & $1^-$ & $1/2$, $3/2$ & $0$ & $0$ & & $1/2$ & $-1$ & & X & X & X \\
$\gamma_0 \gamma_j \gamma_5$ & $1^-$ & $1/2$, $3/2$ & $1$ & $0$ & & $1/2$ & $-1$ & & $0$ & $-2$ & \vspace{-0.40cm} \\
 & & & & & & & & & & & \\
\hline
\end{tabular}
\caption{\label{TAB002}continuum static-light baryon creation operators and their quantum numbers ($j^\mathcal{P}$: angular momentum of the light degrees of freedom and parity; $J$: total angular momentum; $I$: isospin; $S$: strangeness; name: name of the corresponding $b$ baryon(s) in \cite{PDG}); operators marked with ``X'' are identically zero, i.e.\ do not exist.}
\end{center}
\end{table}

% *****

\subsubsection{Static-light baryon creation operators in twisted mass lattice QCD}

Twisted basis lattice static-light baryon creation operators are of similar form,
\begin{eqnarray}
\label{EQN002} \mathcal{O}_{\Gamma,\chi^{(1)} \chi^{(2)}}^\textrm{twisted}(\mathbf{x}) \ \ = \ \ \epsilon^{a b c} Q^a(\mathbf{x}) \Big((\chi^{b,(1)}(\mathbf{x}))^T \mathcal{C} \Gamma \chi^{c,(2)}(\mathbf{x})\Big) ,
\end{eqnarray}
where physical basis quark operators have been replaced by their twisted basis lattice counterparts.

In the continuum the relation between the physical and the twisted basis is given by the twist rotation $\psi = \exp(i \gamma_5 \tau_3 \omega / 2) \chi$, where $\omega = \pi / 2$ at maximal twist. At finite lattice spacing, however, issues are more complicated: the twist rotation only holds for renormalized operators and the QCD symmetries isospin and parity are explicitely broken by $\mathcal{O}(a)$. Nevertheless, it is possible to unambiguously interpret states obtained from correlation functions of twisted basis operators in terms of QCD quantum numbers as we will explain and demonstrate below.

On the lattice rotational symmetry is reduced to symmetry with respect to cubic rotations. There are only five different representations of the cubic group $\mathrm{O}_\mathrm{h}$ corresponding to integer angular momentum $j$. $j = 0$ in the continuum corresponds to the $A_1$ representation on the lattice containing angular momenta $j = 0, 4, 7, \ldots$, while $j = 1$ corresponds to the $T_1$ representation containing $j = 1, 3, 4, \ldots$

While in twisted mass lattice QCD the $z$-component of isospin $I_z$ is still a quantum number, isospin $I$ and parity $\mathcal{P}$ are explicitely broken by the Wilson term, which is proportional to the lattice spacing. Only a specific combination of both symmetries, light flavor exchange combined with parity, is still a symmetry in twisted mass lattice QCD. We denote this symmetry by $\mathcal{P}^{(\textrm{tm})}$ acting on the light twisted basis quark doublet $\chi = (\chi^{(u)} , \chi^{(d)})$ according to $\mathcal{P}^{(\textrm{tm})} \chi = \gamma_0 \tau_1 \chi$, where $\tau_1$ is the first Pauli matrix acting in flavor space. Consequently, the four QCD sectors labeled by $I = 0, 1$ and $\mathcal{P} = +, -$ are pairwise combined. $\mathcal{P}^{(\textrm{tm})} = +$ is a combination of $(I=0,\mathcal{P}=-)$ and $(I=1,\mathcal{P}=+)$, while $\mathcal{P}^{(\textrm{tm})} = -$ is a combination of $(I=0,\mathcal{P}=+)$ and $(I=1,\mathcal{P}=-)$.

As explained in section~\ref{SEC486} the partially quenched $s$ quark can be realized in two ways denoted by $\chi^{(s^+)}$ and $\chi^{(s-)}$, respectively. As a consequence baryons computed at finite lattice spacing on the one hand with $s^+$ quarks and on the other hand with $s^-$ quarks, but which are otherwise identical, may differ in mass. Due to automatic $\mathcal{O}(a)$ improvement of twisted mass lattice QCD this mass splitting, however, will only be $\mathcal{O}(a^2)$, i.e.\ is expected to be rather small and will vanish quadratically, when approaching the continuum limit.

Since $\mathcal{P}^{(\textrm{tm})}$ and $I_z$ do not commute, they cannot simultaneously be chosen as quantum numbers. An exception are states with $I_z = 0$, which can also be classified with respect to $\mathcal{P}^{(\textrm{tm})}$.

The lattice static-light baryon creation operators we have been using are collected in Table~\ref{TAB003}, Table~\ref{TAB004} and Table~\ref{TAB005}. Creation operators are sorted according to the twisted mass lattice quantum numbers of their associated trial states, i.e.\ creation operators exciting states from different sectors are separated by horizontal lines. To interpret these twisted basis creation operators in terms of QCD quantum numbers, we have performed an approximate rotation to the physical basis (neglecting renormalization and using $\omega = \pi / 2$). The resulting so-called pseudo physical basis creation operators together with their corresponding QCD quantum numbers are also listed in the tables.

% **********
% **********
% **********
% S = 0 operators
% **********
% **********
% **********

\begin{table}[htb]
\begin{center}
\begin{tabular}{|c|c||c|c|||c|c||c|c|c||c|}
\hline
\multicolumn{4}{|c|||}{\vspace{-0.40cm}} & \multicolumn{6}{c|}{} \\
\multicolumn{4}{|c|||}{twisted basis lattice operator} & \multicolumn{6}{c|}{pseudo physical basis operator} \\
\multicolumn{4}{|c|||}{\vspace{-0.40cm}} & \multicolumn{6}{c|}{} \\
\hline
 & & & & & & & & & \vspace{-0.40cm} \\
$\Gamma$ & $\chi^{(1)} \chi^{(2)}$ & $I_z$ & $\mathcal{P}^{(\textrm{tm})}$ & $\Gamma$ & $\psi^{(1)} \psi^{(2)}$ & $I$ & $I_z$ & $\mathcal{P}$ & name \vspace{-0.40cm} \\
 & & & & & & & & & \\
\hline
\multicolumn{10}{|c|}{\vspace{-0.40cm}} \\
%
%
% **********
%
%
\multicolumn{10}{|c|}{$A_1 \textrm{ representation} \quad \equiv \quad j = 0, 4, 7, \ldots$} \\
\multicolumn{10}{|c|}{\vspace{-0.40cm}} \\
\hline
 & & & & & & & & & \vspace{-0.40cm} \\
$\gamma_5$          & $ud - du$ & $0$     & $-$   &   $\gamma_5$          & $ud - du$ & $0$ & $0$     & $+$ & $\Lambda_b$ \\
$\gamma_0$          & $ud + du$ & $0$     & $-$   &   $\gamma_0 \gamma_5$ & $ud - du$ & $0$ & $0$     & $+$ & $\Lambda_b$ \\
$\gamma_0 \gamma_5$ & $ud - du$ & $0$     & $-$   &   $\gamma_0$          & $ud + du$ & $1$ & $0$     & $-$ & \\
 & & & & & & & & & \vspace{-0.40cm} \\
\hline
 & & & & & & & & & \vspace{-0.40cm} \\
$1$                 & $ud - du$ & $0$     & $+$   &   $1$                 & $ud - du$ & $0$ & $0$     & $-$ & \\
 & & & & & & & & & \vspace{-0.40cm} \\
\hline
 & & & & & & & & & \vspace{-0.40cm} \\
$\gamma_0$          & $uu$/$dd$ & $+1/-1$ & xxx   &   $\gamma_0$          & $uu$/$dd$ & $1$ & $+1/-1$ & $-$ & \vspace{-0.40cm} \\
 & & & & & & & & & \\
\hline
\multicolumn{10}{|c|}{\vspace{-0.40cm}} \\
%
%
% **********
%
%
\multicolumn{10}{|c|}{$T_1 \textrm{ representation} \quad \equiv \quad j = 1, 3, 4, \ldots$} \\
\multicolumn{10}{|c|}{\vspace{-0.40cm}} \\
\hline
 & & & & & & & & & \vspace{-0.40cm} \\
$\gamma_j \gamma_5$          & $ud - du$ & $0$     & $+$   &   $\gamma_j$                   & $ud + du$ & $1$ & $0$     & $+$ & $\Sigma_b$, $\Sigma_b^\ast$ \\
$\gamma_0 \gamma_j$          & $ud + du$ & $0$     & $+$   &   $\gamma_0 \gamma_j$          & $ud + du$ & $1$ & $0$     & $+$ & $\Sigma_b$, $\Sigma_b^\ast$ \\
$\gamma_j$                   & $ud + du$ & $0$     & $+$   &   $\gamma_j \gamma_5$          & $ud - du$ & $0$ & $0$     & $-$ & \\
 & & & & & & & & & \vspace{-0.40cm} \\
\hline
 & & & & & & & & & \vspace{-0.40cm} \\
$\gamma_0 \gamma_j \gamma_5$ & $ud + du$ & $0$     & $-$   &   $\gamma_0 \gamma_j \gamma_5$ & $ud + du$ & $1$ & $0$     & $-$ & \\
 & & & & & & & & & \vspace{-0.40cm} \\
\hline
 & & & & & & & & & \vspace{-0.40cm} \\
$\gamma_j$                   & $uu$/$dd$ & $+1/-1$ & xxx   &   $\gamma_j$                   & $uu$/$dd$ & $1$ & $+1/-1$ & $+$ & $\Sigma_b$, $\Sigma_b^\ast$ \\
$\gamma_0 \gamma_j \gamma_5$ & $uu$/$dd$ & $+1/-1$ & xxx   &   $\gamma_0 \gamma_j$          & $uu$/$dd$ & $1$ & $+1/-1$ & $+$ & $\Sigma_b$, $\Sigma_b^\ast$ \\
$\gamma_0 \gamma_j$          & $uu$/$dd$ & $+1/-1$ & xxx   &   $\gamma_0 \gamma_j \gamma_5$ & $uu$/$dd$ & $1$ & $+1/-1$ & $-$ & \vspace{-0.40cm} \\
 & & & & & & & & & \\
\hline
\end{tabular}
\caption{\label{TAB003}$S = 0$ lattice static-light baryon creation operators and their quantum numbers; ($j$: angular momentum of the light degrees of freedom; $I$: isospin; $I_z$: $z$-component of isospin; $\mathcal{P}$: parity; $\mathcal{P}^{(\textrm{tm})}$: twisted mass parity [``xxx'' indicates that $\mathcal{P}^{(\textrm{tm})}$ is not a quantum number for the corresponding trial state]; name: name of the corresponding $b$ baryon(s) in \cite{PDG}).}
\end{center}
\end{table}

% **********
% **********
% **********
% S = -1 operators
% **********
% **********
% **********

\begin{table}[htb]
\begin{center}
\begin{tabular}{|c|c||c|||c|c||c|c|c||c|}
\hline
\multicolumn{3}{|c|||}{\vspace{-0.40cm}} & \multicolumn{6}{c|}{} \\
\multicolumn{3}{|c|||}{twisted basis lattice operator} & \multicolumn{6}{c|}{pseudo physical basis operator} \\
\multicolumn{3}{|c|||}{\vspace{-0.40cm}} & \multicolumn{6}{c|}{} \\
\hline
 & & & & & & & & \vspace{-0.40cm} \\
$\Gamma$ & $\chi^{(1)} \chi^{(2)}$ & $I_z$ & $\Gamma$ & $\psi^{(1)} \psi^{(2)}$ & $I$ & $I_z$ & $\mathcal{P}$ & name \vspace{-0.40cm} \\
 & & & & & & & & \\
\hline
\multicolumn{9}{|c|}{\vspace{-0.40cm}} \\
%
%
% **********
%
%
\multicolumn{9}{|c|}{$A_1 \textrm{ representation} \quad \equiv \quad j = 0, 4, 7, \ldots$} \\
\multicolumn{9}{|c|}{\vspace{-0.40cm}} \\
\hline
 & & & & & & & & \vspace{-0.40cm} \\
$1$                 & $us^+$/$ds^-$ & $+1/2/-1/2$   &   $\gamma_5$          & $us$/$ds$ & $1/2$ & $+1/2/-1/2$ & $+$ & $\Xi_b$ \\
$\gamma_0 \gamma_5$ & $us^+$/$ds^-$ & $+1/2/-1/2$   &   $\gamma_0 \gamma_5$ & $us$/$ds$ & $1/2$ & $+1/2/-1/2$ & $+$ & $\Xi_b$ \\
$\gamma_5$          & $us^+$/$ds^-$ & $+1/2/-1/2$   &   $1$                 & $us$/$ds$ & $1/2$ & $+1/2/-1/2$ & $-$ & \\
$\gamma_0$          & $us^+$/$ds^-$ & $+1/2/-1/2$   &   $\gamma_0$          & $us$/$ds$ & $1/2$ & $+1/2/-1/2$ & $-$ & \\
 & & & & & & & & \vspace{-0.40cm} \\
\hline
 & & & & & & & & \vspace{-0.40cm} \\
$\gamma_5$          & $us^-$/$ds^+$ & $+1/2/-1/2$   &   $\gamma_5$          & $us$/$ds$ & $1/2$ & $+1/2/-1/2$ & $+$ & $\Xi_b$ \\
$\gamma_0$          & $us^-$/$ds^+$ & $+1/2/-1/2$   &   $\gamma_0 \gamma_5$ & $us$/$ds$ & $1/2$ & $+1/2/-1/2$ & $+$ & $\Xi_b$ \\
$1$                 & $us^-$/$ds^+$ & $+1/2/-1/2$   &   $1$                 & $us$/$ds$ & $1/2$ & $+1/2/-1/2$ & $-$ & \\
$\gamma_0 \gamma_5$ & $us^-$/$ds^+$ & $+1/2/-1/2$   &   $\gamma_0$          & $us$/$ds$ & $1/2$ & $+1/2/-1/2$ & $-$ & \vspace{-0.40cm} \\
 & & & & & & & & \\
\hline
\multicolumn{9}{|c|}{\vspace{-0.40cm}} \\
%
%
% **********
%
%
\multicolumn{9}{|c|}{$T_1 \textrm{ representation} \quad \equiv \quad j = 1, 3, 4, \ldots$} \\
\multicolumn{9}{|c|}{\vspace{-0.40cm}} \\
\hline
 & & & & & & & & \vspace{-0.40cm} \\
$\gamma_j$                   & $us^+$/$ds^-$ & $+1/2/-1/2$   &   $\gamma_j$                   & $us$/$ds$ & $1/2$ & $+1/2/-1/2$ & $+$ & \\
$\gamma_0 \gamma_j \gamma_5$ & $us^+$/$ds^-$ & $+1/2/-1/2$   &   $\gamma_0 \gamma_j$          & $us$/$ds$ & $1/2$ & $+1/2/-1/2$ & $+$ & \\
$\gamma_j \gamma_5$          & $us^+$/$ds^-$ & $+1/2/-1/2$   &   $\gamma_j \gamma_5$          & $us$/$ds$ & $1/2$ & $+1/2/-1/2$ & $-$ & \\
$\gamma_0 \gamma_j$          & $us^+$/$ds^-$ & $+1/2/-1/2$   &   $\gamma_0 \gamma_j \gamma_5$ & $us$/$ds$ & $1/2$ & $+1/2/-1/2$ & $-$ & \\
 & & & & & & & & \vspace{-0.40cm} \\
\hline
 & & & & & & & & \vspace{-0.40cm} \\
$\gamma_j \gamma_5$          & $us^-$/$ds^+$ & $+1/2/-1/2$   &   $\gamma_j$                   & $us$/$ds$ & $1/2$ & $+1/2/-1/2$ & $+$ & \\
$\gamma_0 \gamma_j$          & $us^-$/$ds^+$ & $+1/2/-1/2$   &   $\gamma_0 \gamma_j$          & $us$/$ds$ & $1/2$ & $+1/2/-1/2$ & $+$ & \\
$\gamma_j$                   & $us^-$/$ds^+$ & $+1/2/-1/2$   &   $\gamma_j \gamma_5$          & $us$/$ds$ & $1/2$ & $+1/2/-1/2$ & $-$ & \\
$\gamma_0 \gamma_j \gamma_5$ & $us^-$/$ds^+$ & $+1/2/-1/2$   &   $\gamma_0 \gamma_j \gamma_5$ & $us$/$ds$ & $1/2$ & $+1/2/-1/2$ & $-$ & \vspace{-0.40cm} \\
 & & & & & & & & \\
\hline
\end{tabular}
\caption{\label{TAB004}$S = -1$ lattice static-light baryon creation operators and their quantum numbers; ($j$: angular momentum of the light degrees of freedom; $I$: isospin; $I_z$: $z$-component of isospin; $\mathcal{P}$: parity; name: name of the corresponding $b$ baryon in \cite{PDG}).}
\end{center}
\end{table}

% **********
% **********
% **********
% S = -2 operators
% **********
% **********
% **********

\begin{table}[htb]
\begin{center}
\begin{tabular}{|c|c||c|||c|c||c|c|c||c|}
\hline
\multicolumn{3}{|c|||}{\vspace{-0.40cm}} & \multicolumn{6}{c|}{} \\
\multicolumn{3}{|c|||}{twisted basis lattice operator} & \multicolumn{6}{c|}{pseudo physical basis operator} \\
\multicolumn{3}{|c|||}{\vspace{-0.40cm}} & \multicolumn{6}{c|}{} \\
\hline
 & & & & & & & & \vspace{-0.40cm} \\
$\Gamma$ & $\chi^{(1)} \chi^{(2)}$ & $I_z$ & $\Gamma$ & $\psi^{(1)} \psi^{(2)}$ & $I$ & $I_z$ & $\mathcal{P}$ & name \vspace{-0.40cm} \\
 & & & & & & & & \\
\hline
\multicolumn{9}{|c|}{\vspace{-0.40cm}} \\
%
%
% **********
%
%
\multicolumn{9}{|c|}{$A_1 \textrm{ representation} \quad \equiv \quad j = 0, 4, 7, \ldots$} \\
\multicolumn{9}{|c|}{\vspace{-0.40cm}} \\
\hline
 & & & & & & & & \vspace{-0.40cm} \\
$\gamma_0$ & $s^+s^+$/$s^-s^-$ & $0$   &   $\gamma_0$          & $ss$ & $0$ & $0$ & $-$ & \vspace{-0.40cm} \\
 & & & & & & & & \\
\hline
\multicolumn{9}{|c|}{\vspace{-0.40cm}} \\
%
%
% **********
%
%
\multicolumn{9}{|c|}{$T_1 \textrm{ representation} \quad \equiv \quad j = 1, 3, 4, \ldots$} \\
\multicolumn{9}{|c|}{\vspace{-0.40cm}} \\
\hline
 & & & & & & & & \vspace{-0.40cm} \\
$\gamma_j$                   & $s^+s^+$/$s^-s^-$ & $0$   &   $\gamma_j$                   & $ss$ & $0$ & $0$ & $+$ & $\Omega_b$ \\
$\gamma_0 \gamma_j \gamma_5$ & $s^+s^+$/$s^-s^-$ & $0$   &   $\gamma_0 \gamma_j$          & $ss$ & $0$ & $0$ & $+$ & $\Omega_b$ \\
$\gamma_0 \gamma_j$          & $s^+s^+$/$s^-s^-$ & $0$   &   $\gamma_0 \gamma_j \gamma_5$ & $ss$ & $0$ & $0$ & $-$ & \vspace{-0.40cm} \\
 & & & & & & & & \\
\hline
\end{tabular}
\caption{\label{TAB005}$S = -2$ lattice static-light baryon creation operators and their quantum numbers; ($j$: angular momentum of the light degrees of freedom; $I$: isospin; $I_z$: $z$-component of isospin; $\mathcal{P}$: parity; name: name of the corresponding $b$ baryon in \cite{PDG}).}
\end{center}
\end{table}

% *****

\subsubsection{Smearing of gauge links and quark fields}

To enhance the overlap of the trial states $\mathcal{O}_{\Gamma,\chi^{(1)} \chi^{(2)}}^\textrm{twisted} | \Omega \rangle$ to low lying static-light baryon states, we make extensive use of standard smearing techniques. This allows to read off static-light baryon masses from correlation functions at rather small temporal separation, where the signal-to-noise ratio is favorable.

% > sqrt(2 * 10 * 0.5 / (1 + 6 * 0.5))
% [1] 1.581139
% > sqrt(2 * 40 * 0.5 / (1 + 6 * 0.5))
% [1] 3.162278
% > sqrt(2 * 90 * 0.5 / (1 + 6 * 0.5))
% [1] 4.743416

% > sqrt(2 * 10 * 0.5 / (1 + 6 * 0.5)) * 0.079
% [1] 0.1249100
% > sqrt(2 * 40 * 0.5 / (1 + 6 * 0.5)) * 0.079
% [1] 0.2498199
% > sqrt(2 * 90 * 0.5 / (1 + 6 * 0.5)) * 0.079
% [1] 0.3747299

Smearing is done in two steps. At first we replace all spatial gauge links by APE smeared versions. The parameters we have chosen are $N_\textrm{APE} = 40$ and $\alpha_\textrm{APE} = 0.5$. Then we use Gaussian smearing on the light quark fields $\chi^{(u)}$, $\chi^{(d)}$, $\chi^{(s^+)}$ and $\chi^{(s^-)}$, which resorts to the APE smeared spatial links. We consider three different smearing levels, characterized by $N_\textrm{Gauss} \in \{ 10 \, , \, 40 \, , \, 90 \}$ and $\kappa_\textrm{Gauss} = 0.5$. This amounts to light quark field operators with approximate widths of \\ $\{ 1.58 \times a \, , \, 3.16 \times a \, , \, 4.74 \times a \} \approx \{ 0.12 \, \textrm{fm} \, , \, 0.25 \, \textrm{fm} \, , \, 0.37 \, \textrm{fm} \}$ (cf.\ \cite{Jansen:2008si} for details).

Smeared static light baryon creation operators are denoted by $S^{N_{\textrm{Gauss}}}(\mathcal{O}_{\Gamma,\chi^{(1)} \chi^{(2)}}^\textrm{twisted})$.

% **********

\subsection{Correlation matrices}

For each sector characterized by strangeness $S$, angular momentum of the light degrees of freedom $j$, $z$-component of isospin $I_z$, and in certain cases twisted mass parity $\mathcal{P}^{(\textrm{tm})}$ we compute temporal correlation matrices
\begin{eqnarray}
\nonumber & & \hspace{-0.7cm} C_{(\Gamma_j,(\chi^{(1)} \chi^{(2)})_j,N_{\textrm{Gauss},j}) , (\Gamma_k,(\chi^{(1)} \chi^{(2)})_k,N_{\textrm{Gauss},k})}(t) \ \ = \\
\label{EQN541} & & = \ \ \langle \Omega | \Big(S^{N_{\textrm{Gauss,j}}}(\mathcal{O}_{\Gamma_j,(\chi^{(1)} \chi^{(2)})_j}^\textrm{twisted}(t))\Big)^\dagger S^{N_{\textrm{Gauss,k}}}(\mathcal{O}_{\Gamma_k,(\chi^{(1)} \chi^{(2)})_k}^\textrm{twisted}(0)) | \Omega \rangle .
\end{eqnarray}
We consider all the creation operators listed in Table~\ref{TAB003}, Table~\ref{TAB004} and Table~\ref{TAB005} at three different smearing levels $N_\textrm{Gauss} \in \{ 10 \, , \, 40 \, , \, 90 \}$ as explained in the previous subsection. This amounts dependent on the sector to $3 \times 3$, $9 \times 9$ or $12 \times 12$ correlation matrices.

Static quarks are treated with the HYP2 static action \cite{Hasenfratz:2001hp,DellaMorte:2003mn,Della Morte:2005yc}, i.e.\ Wilson lines appearing in static quark propagators are formed by products of HYP2 smeared temporal links (cf.\ \cite{Jansen:2008si} for details).

Light quark propagators are estimated by means of $\mathcal{Z}_2 \times \mathcal{Z}_2$ stochastic timeslice sources (cf.\ \cite{Jansen:2008si} for details). On each gauge field configuration we invert 48 independently chosen sources, all located on the same timeslice, 12 for each of the four possible light quark propagators $u$, $d$, $s^+$ and $s^-$. Multiple inversions of the same timeslice of the same gauge field configuration are beneficial with respect to statistical precision, because each correlation function contains two light quark propagators. This allows to form $12 \times 12 = 144$ statistical samples, i.e.\ the number of samples is the square of the number of inversions (cf.\ \cite{Michael:1998sg}).

% **********

\subsection{\label{SEC394}Determination of static-light baryon masses}

\begin{figure}[htb]
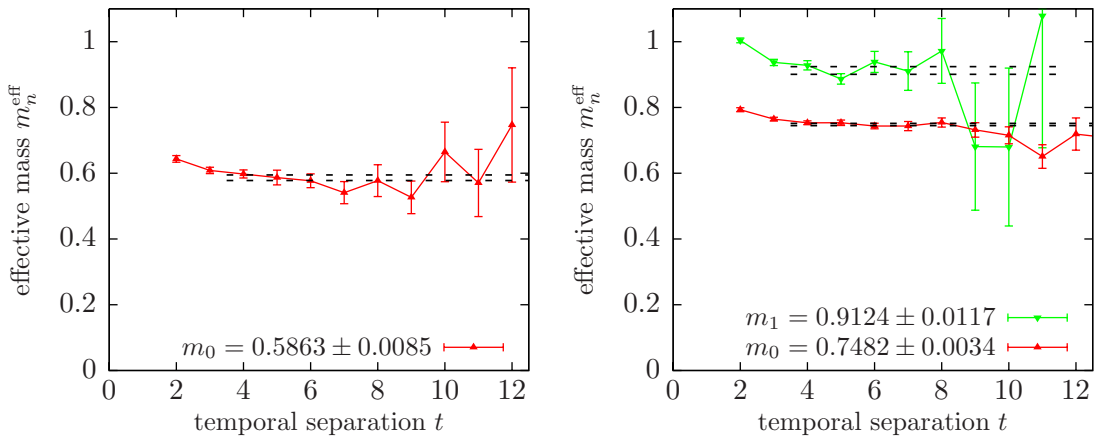

\begin{center}
\input{ud_3x3.tex}\input{s+s+_3x3.tex}
\caption{\label{FIG001}effective masses $m_n^\textrm{eff}$ as functions of the temporal separation $t$ at light quark mass $\mu_\mathrm{q} = 0.0040$; {\bf left}: $\Lambda_b$ ($S=0$, $I=0$, $j^\mathcal{P}=0^+$) from a $3 \times 3$ correlation matrix; {\bf right}: $\Omega_b$ ($S=-2$, $I=0$, $j^\mathcal{P}=1^+$) and its parity partner ($S=-2$, $I=0$, $j^\mathcal{P}=1^-$) from a $3 \times 3$ correlation matrix.}
\end{center}
\end{figure}

From correlation matrices (\ref{EQN541}) we compute effective mass plateaus by solving generalized eigenvalue problems
\begin{eqnarray}
C(t) v_n(t,t_0) \ \ = \ \ \lambda_n(t,t_0) C(t_0) v_n(t,t_0) \quad , \quad m_n^\textrm{eff}(t,t_0) \ \ = \ \ \ln \frac{\lambda_n(t,t_0)}{\lambda_n(t+1,t_0)}
\end{eqnarray}
with $t_0 = 1$ (cf.\ e.g.\ \cite{Michael:1985ne,Blossier:2009kd}). Instead of using the full $3 \times 3$, $9 \times 9$ or $12 \times 12$ correlation matrices we have chosen ``optimal submatrices'' in a sense that on the one hand effective masses exhibit plateaus already at small temporal separations $t$ and that on the other hand statistical errors on $m_n^\textrm{eff}$ are minimized. We found that with the following choice both criteria are adequately fulfilled:
\begin{itemize}
\item $3 \times 3$ correlation matrices: \\
use $2 \times 2$ submatrices with smearing levels $N_\textrm{Gauss} \in \{ 40 \, , \, 90 \}$;

\item $9 \times 9$ correlation matrices: \\
use $3 \times 3$ submatrices with smearing levels $N_\textrm{Gauss} = 90$;

\item $12 \times 12$ correlation matrices: \\
use $4 \times 4$ submatrices with smearing levels $N_\textrm{Gauss} = 90$.
\end{itemize}
To demonstrate the quality of our lattice results, we show in Figure~\ref{FIG001} examples of effective mass plateaus (at light quark mass $\mu_\mathrm{q} = 0.0040$) corresponding to $\Lambda_b$ ($S=0$, $I=0$, $j^\mathcal{P}=0^+$), $\Omega_b$ ($S=-2$, $I=0$, $j^\mathcal{P}=1^+$) and its parity partner ($S=-2$, $I=0$, $j^\mathcal{P}=1^-$).

We extract static-light baryon masses by fitting constants to these plateaus in regions of sufficiently large temporal separation $t_\textrm{min} \ldots t_\textrm{max}$. We found that $t_\textrm{min} = 4$ yields reasonable $\chi^2$ values, which are $\mathcal{O}(1)$ for all states investigated. $t_\textrm{max}$ on the other hand hardly affects the resulting static-light baryon masses (on the ``$t_\textrm{max}$-side'' of the effective mass plateau statistical errors are rather large and, therefore, data points have a negligible effect on the fit). The resulting fits for the examples shown in Figure~\ref{FIG001} are indicated by dashed lines. We checked the stability of all our results by varying $t_\textrm{min}$ by $\pm 1$. We found consistency within statistical errors.

\begin{figure}[htb]
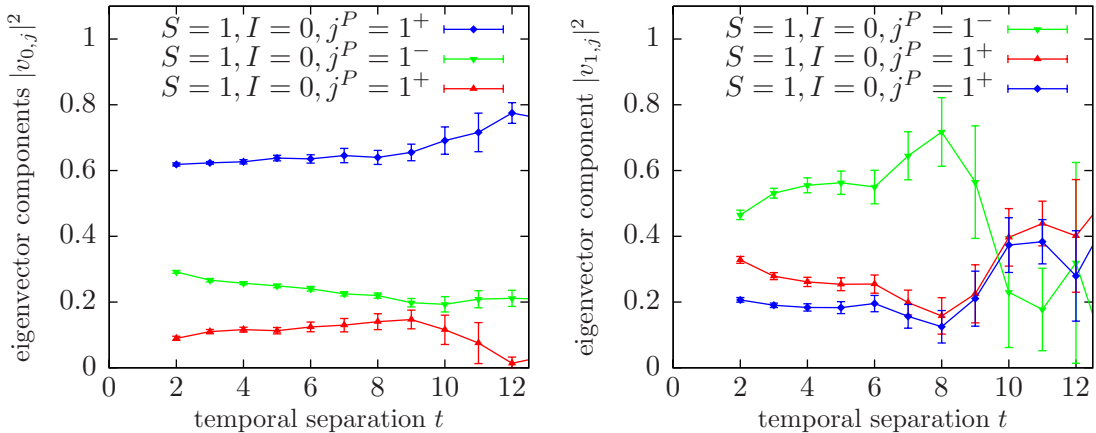

\begin{center}
\input{EV_s+s+_0.tex}\input{EV_s+s+_1.tex}
\caption{\label{FIG002}eigenvector components $|v_{n,j}|^2$ as functions of the temporal separation $t$ and their associated QCD quantum numbers at light quark mass $\mu_\mathrm{q} = 0.0040$ corresponding to the $3 \times 3$ correlation matrix with twisted mass quantum numbers ($S=-2$, $j=1$, $I=0$) (cf.\ also Figure~\ref{FIG001}); {\bf left}: ground state identified as $\Omega_b$ ($S=-2$, $I=0$, $j^\mathcal{P}=1^+$); {\bf right}: first excited state identified as parity partner of $\Omega_b$ ($S=-2$, $I=0$, $j^\mathcal{P}=1^-$).}
\end{center}
\end{figure}

To assign appropriate QCD quantum numbers to the extracted static-light baryon states, we follow a method introduced and explained in detail in \cite{Baron:2010th}, section~3.1 (``Method 1: solving a generalized eigenvalue problem''). For the $n$-th state the components of the corresponding eigenvector $v_{n,j}$ characterize the contribution of the $j$-th static-light baryon creation operator entering the correlation matrix. After transforming these operators from the twisted basis to the pseudo physical basis by means of the twist rotation $\psi = \exp(i \gamma_5 \tau_3 \omega / 2) \chi$, $\omega = \pi / 2$ (cf.\ the right columns of Table~\ref{TAB003}, Table~\ref{TAB004} and Table~\ref{TAB005}), one expects and and also finds that for each extracted state operators corresponding to only one of the two QCD sectors corresponding to the investigated twisted mass lattice QCD sector clearly dominate, while the contribution from operators from the other sector are rather small. This allows to unambiguously assign a QCD label to each extracted static-light baryon state. An example, the identification of $\Omega_b$ ($S=-2$, $I=0$, $j^\mathcal{P}=1^+$) and its parity partner ($S=-2$, $I=0$, $j^\mathcal{P}=1^-$), is shown in Figure~\ref{FIG002} (cf.\ also Figure~\ref{FIG001} for the corresponding effective masses both having twisted mass quantum numbers ($S=-2$, $j=1$, $I=0$)).

Since static-light baryon masses diverge in the continuum limit due to the self energy of the static quark, we always consider mass differences of these baryons to the lightest static-light meson (``$B/B^\ast$ meson''). In such differences the divergent self energy exactly cancels. We take the mass values of the lightest static-light mesons from \cite{:2010iv}, where they have been computed using the same lattice setup. The mass differences $\Delta m^\textrm{stat}(S,I,j^\mathcal{P}) a = (m(\textrm{baryon}:S,I,j^\mathcal{P}) - m(B/B^\ast)) a$ (in lattice units) together with the pion masses $m_\textrm{PS} a$ (also in lattice units; cf.\ Table~\ref{TAB100} and \cite{Baron:2009wt}) serve as input for the extrapolation procedure to the physical $u/d$ quark mass described in the next subsection.

% \mu = 0.0040
% No indication of a systematical error present!
% state 0: m = 0.397484 +/- 0.002183 +/- 0.000000 ( 5 <= t <= 11, \chi^2/dof =  1.42)
% --> state 0: m = 0.397484 +/- 0.002183 ( 5 <= t <= 11, \chi^2/dof =  1.42)

% \mu = 0.0064
% No indication of a systematical error present!
% state 0: m = 0.403754 +/- 0.001750 +/- 0.000000 ( 5 <= t <= 11, \chi^2/dof =  1.46)
% --> state 0: m = 0.403754 +/- 0.001750 ( 5 <= t <= 11, \chi^2/dof =  1.46)

% \mu = 0.0085
% No indication of a systematical error present!
% state 0: m = 0.411901 +/- 0.001211 +/- 0.000000 ( 4 <= t <= 11, \chi^2/dof =  1.95)
% --> state 0: m = 0.411901 +/- 0.001211 ( 4 <= t <= 11, \chi^2/dof =  1.95)
% ##### 0.411490 +/- 0.001362 (for T_min = 5)

% \mu = 0.0100
% No indication of a systematical error present!
% state 0: m = 0.414343 +/- 0.001426 +/- 0.000000 ( 4 <= t <= 11, \chi^2/dof =  1.22)
% --> state 0: m = 0.414343 +/- 0.001426 ( 4 <= t <= 11, \chi^2/dof =  1.22)
% ##### 0.413089 +/- 0.001571 (for T_min = 5)

% **********

\subsection{\label{SEC461}Extrapolation to the physical $u/d$ quark mass}

\begin{figure}[htp]
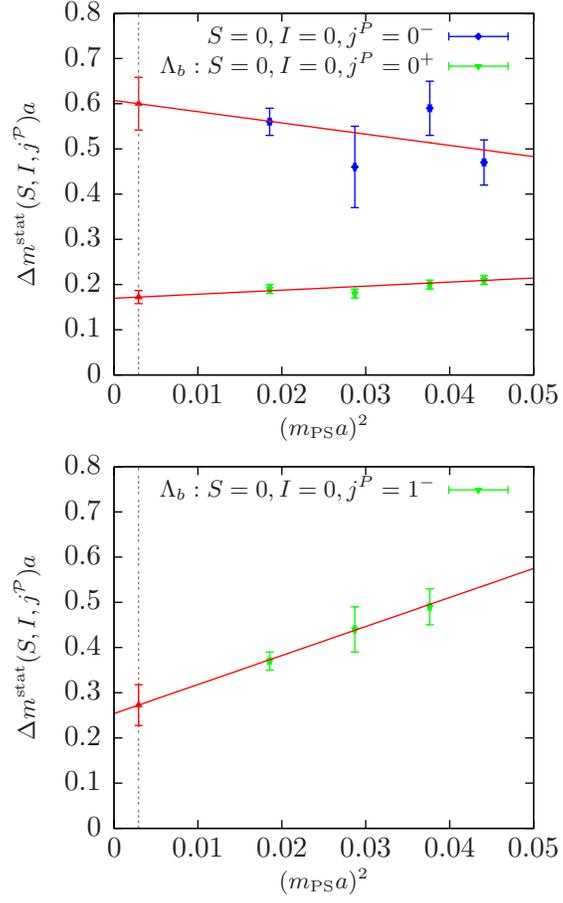

\begin{center}
\input{extra_ud_j=0.tex}\\
\input{extra_ud_i=0_j=1.tex}
\caption{\label{FIG003a}mass differences of $S = 0$, $I = 0$ static-light baryons ($\Lambda$ baryons) to the lightest static-light meson as functions of $(m_\textrm{PS} a)^2$; straight lines represent linear extrapolations to the physical $u/d$ quark mass.}
\end{center}
\end{figure}

\begin{figure}[htp]
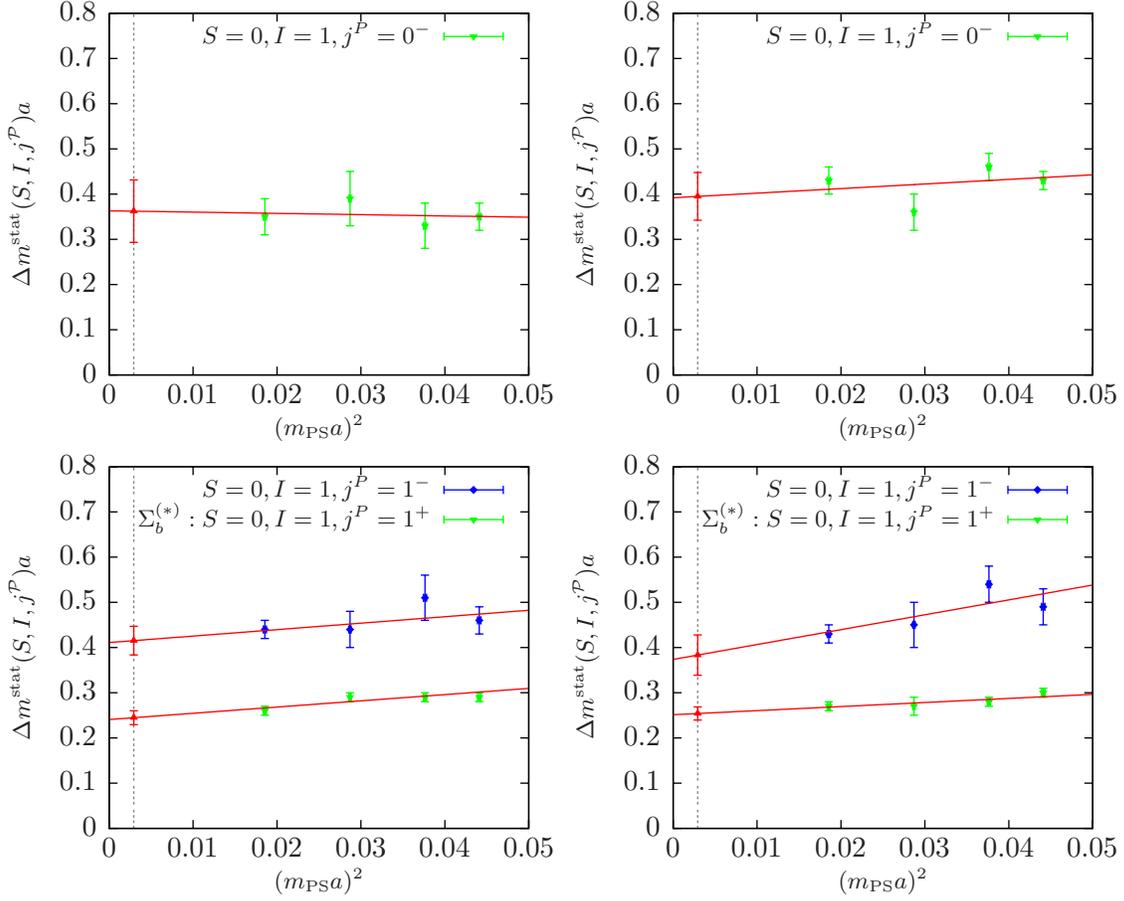

\begin{center}
\input{extra_ud_i=1_j=0.tex}\input{extra_uu_j=0.tex}
\input{extra_ud_j=1.tex}\input{extra_uu_j=1.tex}
\caption{\label{FIG003b}mass differences of $S = 0$, $I = 1$ static-light baryons ($\Sigma$ baryons) to the lightest static-light meson as functions of $(m_\textrm{PS} a)^2$; straight lines represent linear extrapolations to the physical $u/d$ quark mass; plots in the same line only differ in $I_z$ (\textbf{left}: $I_z = 0$; \textbf{right}: $I_z = \pm 1$).}
\end{center}
\end{figure}

The mass differences $\Delta m^\textrm{stat}(S,I,j^\mathcal{P}) a = (m(\textrm{baryon}:S,I,j^\mathcal{P}) - m(B/B^\ast)) a$ obtained for the four ensembles listed in Table~\ref{TAB100}, which only differ in the value of the $u/d$ quark mass (both sea and valence), are plotted against $(m_\textrm{PS} a)^2$ in Figure~\ref{FIG003a} ($S = 0$, $I = 0$, i.e.\ $\Lambda$ baryons), Figure~\ref{FIG003b} ($S = 0$, $I = 1$, i.e.\ $\Sigma$ baryons), Figure~\ref{FIG004} ($S = -1$, i.e.\ $\Xi$ baryons) and Figure~\ref{FIG005} ($S = -2$, i.e.\ $\Omega$ baryons) and are collected in appendix~\ref{APP001}.

\begin{figure}[htp]
\begin{center}
\input{extra_s-u_j=0.tex}\input{extra_s+u_j=0.tex}
\input{extra_s-u_j=1.tex}\input{extra_s+u_j=1.tex}
\caption{\label{FIG004}mass differences of $S = -1$ static-light baryons ($\Xi$ baryons) to the lightest static-light meson as functions of $(m_\textrm{PS} a)^2$; straight lines represent linear extrapolations to the physical $u/d$ quark mass; plots in the same line only differ in the sign of the twisted mass term of the $s$ valence quark (\textbf{left}: $u s^+$/$d s^-$; \textbf{right}: $u s^-$/$d s^+$).}
\end{center}
\end{figure}

\begin{figure}[htp]
\begin{center}
\input{extra_s+s+_j=0.tex}\input{extra_s+s-_j=0.tex}\\
\input{extra_s+s+_j=1.tex}\input{extra_s+s-_j=1.tex}
\caption{\label{FIG005}mass differences of $S = -2$ static-light baryons ($\Omega$ baryons) to the lightest static-light meson as functions of $(m_\textrm{PS} a)^2$; straight lines represent linear extrapolations to the physical $u/d$ quark mass; plots in the same line only differ in the signs of the twisted mass terms of the $s$ valence quarks (\textbf{left}: $s^+ s^+$/$s^- s^-$; \textbf{right}: $s^+ s^-$/$s^- s^+$).}
\end{center}
\end{figure}

For the extrapolation to the physical $u/d$ quark mass one could use an effective field theory approach (Chiral HQET for example) as used e.g.\ to study static-light meson decay constants \cite{Blossier:2009gd}. However, this approach has not fully been developed to discuss mass differences $\Delta m^\textrm{stat}(S,I,j^\mathcal{P}) a$ between excited static-light baryon states and the lightest static-light meson so is not appropriate here. Instead we use the simplest assumption, which is supported by our results: a linear dependence in $(m_\textrm{PS} a)^2$.

Data points $((m_\textrm{PS} a)^2 \, , \, \Delta m^\textrm{stat}(S,I,j^\mathcal{P}) a)$ are correlated via $(m_\textrm{PS} a)^2$ in case they correspond to the same ensemble, i.e.\ to the same value of the $u/d$ quark mass. We take that into account via a covariance matrix, which we estimate by resampling $m_\textrm{PS} a$ and all extracted static-light mass differences $\Delta m^\textrm{stat}(S,I,j^\mathcal{P}) a$ ($10 \, 000 \, 000$ samples). Consequently, we do not fit straight lines to the data points $((m_\textrm{PS} a)^2 \, , \, \Delta m^\textrm{stat}(S,I,j^\mathcal{P}) a)$ individually for every static-light baryon state, but perform a single correlated fit of $23$ straight lines to the $23$ mass differences considered. During the fitting we take statistical errors both along the horizontal axis (errors in $m_\textrm{PS} a$) and along the vertical axis (errors in $\Delta m^\textrm{stat}(S,I,j^\mathcal{P}) a$) into account. The method for performing such two-dimensional fits is explained in detail in \cite{:2010iv}.

We find that a fit, which is linear in the light quark mass (represented by the mass squared of the light-light pseudoscalar meson $(m_\textrm{PS} a)^2$) is acceptable, i.e.\ yields $\chi^2 / \textrm{dof} \approx 0.59 \ltapprox 1$. This fit enables us to extrapolate to the physical $u/d$ quark mass, in this work taken as $m_\textrm{PS} = 135 \, \textrm{MeV}$ and converted to lattice units by using the lattice spacing $a = 0.079 \, \textrm{fm}$ \cite{Baron:2009wt} resulting in $(m_\textrm{PS} a)^2 = 0.054^2$ (cf.\ Figure~\ref{FIG003a}, Figure~\ref{FIG003b}, Figure~\ref{FIG004} and Figure~\ref{FIG005}).

Extrapolations of static-light mass differences to the physical $u/d$ quark mass are listed in Table~\ref{TAB191} in $\textrm{MeV}$. Since there seems to be a controversy of around 10\% regarding the value of the lattice spacing in physical units, when using on the one hand the pion mass $m_\pi$ and the pion decay constant $f_\pi$ and on the other hand the pion mass $m_\pi$ and the nucleon mass $m_N$ to set the scale ($a = 0.079(3) \, \textrm{fm}$ \cite{Baron:2009wt} versus $a = 0.089(5) \, \textrm{fm}$ \cite{Alexandrou:2011db}), we also list dimensionless ratios of static-light mass differences,
\begin{eqnarray}
\label{EQN693} R^\textrm{stat}(S,I,j^\mathcal{P}) \ \ = \ \ \frac{\Delta m^\textrm{stat}(S,I,j^\mathcal{P}) a}{\Delta m^\textrm{stat}(\Omega_b) a} .
\end{eqnarray}
These ratios are pretty independent of the lattice spacing and, therefore, preferable, when making predictions or when comparing to other lattice or model computations or to experimental results.

\begin{table}[htp]
\begin{center}
\begin{tabular}{|c|c|c|c|c||c|c||c|}
\hline
 & & & & & & & \vspace{-0.40cm} \\
    &     &                 &      &        & $\Delta m^\textrm{stat}$ in $\textrm{MeV}$, & $\Delta m^\textrm{stat}$ in $\textrm{MeV}$, &                   \\
$S$ & $I$ & $j^\mathcal{P}$ & name & flavor & $a$ from \cite{Baron:2009wt}                & $a$ from \cite{Alexandrou:2011db}           & $R^\textrm{stat}$ \\
 & & & & & & & \vspace{-0.40cm} \\
\hline
% %%%%%%%%%%
% \Lambda
% %%%%%%%%%%
 & & & & & & & \vspace{-0.40cm} \\
$0$ & $0$ & $0^+$ & $\Lambda_b$               & $ud-du$   & $\phantom{0}430(40)\phantom{0}$ & $\phantom{0}382(39)\phantom{0}$  & $0.480(42)$ \\
 & & & & & & & \vspace{-0.32cm} \\
    &     & $0^-$ &                             & $ud-du$   & $1499(156)$                     & $1330(149)$                      & $1.672(166)$ \\
 & & & & & & & \vspace{-0.32cm} \\
    &     & $1^-$ &                             & $ud-du$   & $\phantom{0}681(116)$           & $\phantom{0}605(106)$            & $0.760(127)$ \\
 & & & & & & & \vspace{-0.40cm} \\
\hline
% %%%%%%%%%%
% \Sigma
% %%%%%%%%%%
 & & & & & & & \vspace{-0.40cm} \\
$0$ & $1$ & $1^+$ & $\Sigma_b$, $\Sigma_b^\ast$ & $ud+du$   & $\phantom{0}611(45)\phantom{0}$ & $\phantom{0}543(46)\phantom{0}$  & $0.682(45)$ \\
    &     &       &                             & $uu$/$dd$ & $\phantom{0}635(44)\phantom{0}$ & $\phantom{0}563(45)\phantom{0}$  &  \\
 & & & & & & & \vspace{-0.32cm} \\
    &     & $0^-$ &                             & $ud+du$   & $\phantom{0}905(176)$           & $\phantom{0}803(160)$            & $1.010(194)$ \\
    &     &       &                             & $uu$/$dd$ & $\phantom{0}986(137)$           & $\phantom{0}876(127)$            &  \\
 & & & & & & & \vspace{-0.32cm} \\
    &     & $1^-$ &                            & $ud+du$   & $1037(88)\phantom{0}$           & $\phantom{0}921(87)\phantom{0}$  & $1.158(91)$ \\
    &     &       &                             & $uu$/$dd$ & $\phantom{0}957(117)$           & $\phantom{0}850(110)$            &  \\
 & & & & & & & \vspace{-0.40cm} \\
\hline
% %%%%%%%%%%
% \Xi
% %%%%%%%%%%
 & & & & & & & \vspace{-0.40cm} \\
$-1$ & $1/2$ & $0^+$ & $\Xi_b$ & $us^-$/$ds^+$ & $\phantom{0}602(33)\phantom{0}$ & $\phantom{0}534(37)\phantom{0}$ & $0.672(30)$ \\
     &       &       &           & $us^+$/$ds^-$ & $\phantom{0}629(41)\phantom{0}$ & $\phantom{0}558(43)\phantom{0}$ &  \\
 & & & & & & & \vspace{-0.32cm} \\
     &       & $1^+$ &           & $us^-$/$ds^+$ & $\phantom{0}767(39)\phantom{0}$ & $\phantom{0}681(44)\phantom{0}$ & $0.856(33)$ \\
     &       &       &           & $us^+$/$ds^-$ & $\phantom{0}778(38)\phantom{0}$ & $\phantom{0}690(44)\phantom{0}$ &  \\
 & & & & & & & \vspace{-0.32cm} \\
     &       & $0^-$ &           & $us^-$/$ds^+$ & $1205(117)$                     & $1070(113)$                     & $1.351(123)$ \\
     &       &       &           &  $us^+$/$ds^-$ & $\phantom{0}954(141)$           & $\phantom{0}847(130)$           &  \\
 & & & & & & & \vspace{-0.32cm} \\
     &       & $1^-$ &           & $us^-$/$ds^+$ & $1062(71)\phantom{0}$           & $\phantom{0}943(74)\phantom{0}$ & $1.185(69)$ \\
     &       &       &           & $us^+$/$ds^-$ & $1068(91)\phantom{0}$           & $\phantom{0}948(89)\phantom{0}$ & \\
 & & & & & & & \vspace{-0.40cm} \\
\hline
% %%%%%%%%%%
% \Omega
% %%%%%%%%%%
 & & & & & & & \vspace{-0.40cm} \\
$-2$ & $0$ & $1^+$ & $\Omega_b$ & $s^+s^+$/$s^-s^-$ & $\phantom{0}896(39)\phantom{0}$ & $\phantom{0}795(48)\phantom{0}$ & $1$ \\
     &     &       &              & $s^+s^-$/$s^-s^+$ & $\phantom{0}896(38)\phantom{0}$ & $\phantom{0}795(47)\phantom{0}$ &   \\
 & & & & & & & \vspace{-0.32cm} \\
     &     & $0^-$ &              & $s^+s^+$/$s^-s^-$ & $1336(64)\phantom{0}$           & $1186(75)\phantom{0}$           & $1.491(53)$ \\
     &     &       &              & $s^+s^-$/$s^-s^+$ & $1296(94)\phantom{0}$           & $1150(96)\phantom{0}$           & \\
 & & & & & & & \vspace{-0.32cm} \\
     &     & $1^-$ &              &$s^+s^+$/$s^-s^-$ & $1236(76)\phantom{0}$           & $1097(81)\phantom{0}$           & $1.380(72)$ \\
     &     &       &              & $s^+s^-$/$s^-s^+$  & $1255(61)\phantom{0}$           & $1114(71)\phantom{0}$           & \vspace{-0.40cm} \\
 & & & & & & & \\
\hline
\end{tabular}
\caption{\label{TAB191}static-light mass differences $\Delta m^\textrm{stat}(S,I,j^\mathcal{P}) = m(\textrm{baryon}:S,I,j^\mathcal{P}) - m(B/B^\ast)$ in $\textrm{MeV}$ (scale setting via $m_\pi$ and $f_\pi$, $a = 0.079(3) \, \textrm{fm}$ \cite{Baron:2009wt}, and via $m_\pi$ and $m_N$, $a = 0.089(5) \, \textrm{fm}$ \cite{Alexandrou:2011db}) and dimensionless ratios of static-light mass differences (cf.\ (\ref{EQN693})).}
\end{center}
\end{table}

For static-light baryons with $S = -1$ and $S = -2$ our results depend on the bare $s$ quark mass chosen. We use $\mu_{\mathrm{q},\textrm{valence }s} = 0.022$ taken from studies of strange-light mesons \cite{Blossier:2007vv,Blossier:2009bx} as mentioned in section~\ref{SEC486}. Possible systematic errors arising from a slightly incorrect value of the $s$ quark mass are expected to be smaller than the corresponding statistical errors, because the mass differences we compute turn out to be rather weakly dependent on the masses of their light valence quarks (cf.\ Figure~\ref{FIG003a}, Figure~\ref{FIG003b} and Figure~\ref{FIG004}). A possibility to estimate their magnitude for $\Omega_b$ is to estimate the slope of $m(\Omega_b)$ as a function of $\mu_{\mathrm{q},\textrm{valence }s}$ by means of the experimental results on $m(\Omega_b)$ and $m(\Sigma_b)$ and assuming a systematic error of 10\% for the lattice spacing (i.e.\ roughly the difference obtained with the two scale setting methods \cite{Baron:2009wt,Alexandrou:2011db}). Then one arrives at a systematic error of around $0.1 \times (m(\Omega_b) - m(\Sigma_b)) \approx 23 \, \textrm{MeV}$. This number is consistent with an even simpler method of estimation namely just taking a systematic error of $0.1 \times m_s$ for every $s$ valence quark, where $m_s = 80 \, \textrm{MeV} \ldots 130 \, \textrm{MeV}$ \cite{PDG}. We intend to investigate the $s$ quark dependence in more detail and to quantify the corresponding systematic error more precisely in a subsequent publication.

% A crude estimate of the systematic error:
% \Omega_b has two s quarks and a rather small statistical error, i.e.
% should be affected more by a possible systematic error than all other states.
% Assume a systematic error of 10% on \mu_s (0709.4574, Table 2).
% The \mu_s dependence can be estimated from \Sigma_b and \Omega_b:
% \Delta_syst = (m(\Omega_b) - m(\Sigma_b)) * 0.1 ~ 25 MeV.

Static-light baryon states with ($S=0$, $I=1$) and either $I_z = 0$ ($ud + du$) or $I_z = \pm 1$ ($uu$/$dd$) are not degenerate in twisted mass lattice QCD, but differ by discretization errors. These discretization errors are, however, only $\mathcal{O}(a^2)$ and, therefore, expected to be rather small. As can be seen from Table~\ref{TAB191} $I_z = 0$ and $I_z = \pm 1$ states agree within statistical errors, which is a strong indication that discretization errors are indeed negligible. For the ratios $R^\textrm{stat}(S,I,j^\mathcal{P})$ and for interpolations to the physical $b$ quark mass (cf.\ section~\ref{SEC941}) we subsequently use $I_z = 0$ results.

For ($S=-1$, $I=1/2$) static-light baryon states similar statements apply.

For static-light baryon states with two $s$ quarks, i.e.\ ($S=-2$, $I=0$), the situation is somewhat different. On the operator level one should not use different lattice discretizations of the two $s$ quarks, i.e.\ one twisted $s^+$ and one twisted $s^-$ quark (for example the operator with $\Gamma = \gamma_5$, $\chi^{(1)} \chi^{(2)} = ss - ss$ clearly is identically zero, while $\Gamma = \gamma_5$, $\chi^{(1)} \chi^{(2)} = s^+s^- - s^-s^+$ would give a non-zero correlator, because there is no propagation from $s^+$ to $s^-$). Only on the level of correlators one can insert one $s^+$ lattice propagator and one $s^-$ lattice propagator in a meaningful way. The corresponding masses are also listed in Table~\ref{TAB191} and agree with their $s^+s^+$/$s^-s^-$ counterparts. For the ratios $R^\textrm{stat}(S,I,j^\mathcal{P})$ and for interpolations to the physical $b$ quark mass (cf.\ section~\ref{SEC941}) we subsequently use $s^+s^+$/$s^-s^-$ results.

In principle contamination of some of the computed static-light baryon masses by multiparticle states (e.g.\ by a lighter static-light baryon and a pion) can at this stage not rigorously be excluded. However, from previous experience with similar lattice computations one strongly expects that the overlap of the used trial states to multiparticle states is extremely small and, therefore, that contamination of the obtained static-light baryon masses is negligible within statistical errors. A possibility to investigate this issue in detail is to compute matrix elements of two operators, where one is a ``single particle baryon operator'' and the other a ``multiparticle operator'' (cf.\ \cite{McNeile:2000xx,McNeile:2002az,McNeile:2004rf}, where the method has been introduced and applied to glueballs and various types of mesons). Using this method we were e.g.\ able to confirm that the parity partner of the lightest static-light meson is essentially not affected by multiparticle states (cf.\ \cite{:2010iv}).

Finally it is interesting to compare our static-light mass differences to recent results obtained by other lattice groups. In \cite{Burch:2008qx} three $b$ baryon states, $\Lambda_b$, $\Sigma_b/\Sigma_b^\ast$ and $\Omega_b$, are computed. The method of scale setting used in this paper, imposing $r_0 = 0.49 \, \textrm{fm}$, can easily be applied to our lattice results, which are then in excellent agreement within statistical errors. Since in \cite{Detmold:2008ww} and \cite{Lin:2009rx,Lin:2010wb} scale setting methods are used, which are less straightforward to adapt (via $m_\pi$, $m_K$ and $m_\Omega$ and chiral perturbation theory \cite{Detmold:2008ww} and via $\Upsilon$ \cite{Lin:2009rx,Lin:2010wb}), we directly compare the resulting mass differences in $\textrm{MeV}$. When comparing to those of our results corresponding to $a = 0.089(5) \, \textrm{fm}$ (i.e.\ scale setting via $m_N$), we also find agreement for all states computed in \cite{Detmold:2008ww,Lin:2009rx,Lin:2010wb}, $\Lambda_b$, $\Sigma_b/\Sigma_b^\ast$, $\Xi_b$, $\Xi'_b$ ($S = -2$, $I = 1/2$, $j^\mathcal{P} = 1^+$) and $\Omega_b$.

We are also able to predict eight static-light baryon states of negative parity, for which no other lattice results seem to be available at the moment. Therefore, we compare these predictions to the quark model calculation in \cite{Ebert:2007nw}. Also here we find agreement within statistical errors.

\newpage

\section{\label{SEC941}Interpolation to the physical $b$ quark mass}

To make contact with experimentally available results on the spectrum of $b$ baryons, we need to correct for the finite mass of the $b$ quark. In Heavy Quark Effective Theory the leading correction is $\mathcal{O}(1/m_Q)$, where $m_Q$ is the mass of the heavy quark. It is possible in principle to compute such corrections from first principles by means of lattice QCD (cf.\ e.g.\ \cite{Blossier:2010jk,Blossier:2010vz}). This we intend to explore in the future, but here we use a more direct method, to establish the size of the correction between static quarks and $b$ quarks of physically realistic mass.

% m(\Sigma_b^\ast) - m(\Sigma_b) = 21.2(2.0)MeV

% m(\Sigma_c(2455)) = 2452.5 ... 2454.2 MeV
% m(\Sigma_c(2520)) = 2515.2 ... 2519.8 MeV
% m(\Sigma_c(2520)) - m(\Sigma_c(2455)) = 61 ... 67.3

% 0.35 * ( m(\Sigma_c(2520)) - m(\Sigma_c(2455)) ) = 21.35 ... 23.56

We linearly interpolate in $m_c/m_Q$ between our static-light lattice results and corresponding experimental data for charmed baryons. As a measure of the heavy quark mass $m_Q$ we take the masses of the ground state heavy-light mesons ($D$ or $B$), i.e.\ we interpolate to $m_c/m_b = m(D) / m(B) = 0.35$. This measure is equivalent to another (such as using quark masses in some scheme) to the order $1/m_Q$ we are considering. One test of this interpolation can be made: the hyperfine splitting between $\Sigma_c(2520)$ and $\Sigma_c(2455)$ is around $64 \, \textrm{MeV}$; interpolating between this number and the static limit results in $0.35 \times 64 \, \textrm{MeV} = 22 \, \textrm{MeV}$, which is in fair agreement with the observed splitting of $m(\Sigma_b^\ast) - m(\Sigma_b) = 21 \, \textrm{MeV}$ \cite{PDG}. Results of these interpolations are collected in Table~\ref{TAB192}. Note that these $m_c/m_Q$ corrections break the heavy spin degeneracy of static-light baryon states with $j = 1$.

\begin{table}[htp]
\begin{center}
\begin{tabular}{|c|c|c|c||c|c|c||c|c|}
\hline
 & & & & & & & & \vspace{-0.40cm} \\
    &     &                 &            & $\Delta m^{\textrm{lat},b}$ & $\Delta m^{\textrm{lat},b}$   & $\Delta m^{\textrm{exp},b}$ &                      &                      \\
    &     &                 &            & in $\textrm{MeV}$, $a$      & in $\textrm{MeV}$, $a$        & in $\textrm{MeV}$ $^1$           &                      &                      \\ 
$S$ & $I$ & $J^\mathcal{P}$ & $b/c$ name & from \cite{Baron:2009wt}    & from \cite{Alexandrou:2011db} &                             & $R^{\textrm{lat},b}$ & $R^{\textrm{exp},b}$ $^1$ \\
 & & & & & & & & \vspace{-0.40cm} \\
\hline
% %%%%%%%%%%
% \Lambda
% %%%%%%%%%%
 & & & & & & & & \vspace{-0.40cm} \\
$0$ & $0$ & $(1/2)^+$ & $\Lambda_b$/$\Lambda_c$ & $ 426(26)$ & $ 395(25)$ & $341(2)$ & $0.489(27)$ & $0.440(5)$ \\
 & & & & & & & & \vspace{-0.32cm} \\
    &     & $(1/2)^-$ & $-$/$-$                     & $-$ & $-$ & $-$ & $-$ & $-$ \\
 & & & & & & & & \vspace{-0.32cm} \\
    &     & $(1/2)^-$ & $-$/$\Lambda_c(2595)$     & $697(75)$ & $648(69)$ & $-$ & $0.802(83)$ & $-$ \\
    &     & $(3/2)^-$ & $-$/$\Lambda_c(2625)$     & $709(75)$ & $660(69)$ & $-$ & $0.816(83)$ & $-$ \\
 & & & & & & & & \vspace{-0.40cm} \\
\hline
% %%%%%%%%%%
% \Sigma
% %%%%%%%%%%
 & & & & & & & & \vspace{-0.40cm} \\
$0$ & $1$ & $(1/2)^+$ & $\Sigma_b$/$\Sigma_c(2455)$      & $602(29)$ & $558(30)$ & 532(6)  &$0.691(30)$ & $0.687(11)$ \\
    &     & $(3/2)^+$ & $\Sigma_b^\ast$/$\Sigma_c(2520)$ & $628(29)$ & $584(30)$ & 553(7) &$0.718(30)$ & $0.714(11)$ \\
 & & & & & & & & \vspace{-0.32cm} \\
    &     & $(1/2)^-$ & $-$/$-$                     & $-$ & $-$ & $-$ & $-$ & $-$ \\
 & & & & & & & & \vspace{-0.32cm} \\
    &     & $(1/2)^-$ & $-$/$-$                     & $-$ & $-$ & $-$ & $-$ & $-$ \\
    &     & $(3/2)^-$ & $-$/$-$                     & $-$ & $-$ & $-$ & $-$ & $-$ \\
 & & & & & & & & \vspace{-0.40cm} \\
\hline
% %%%%%%%%%%
% \Xi
% %%%%%%%%%%
 & & & & & & & & \vspace{-0.40cm} \\
$-1$ & $1/2$ & $(1/2)^+$ & $\Xi_b$/$\Xi_c$ & $602(21)$ & $ 558 (24)$ & 511(3)& $0.691(20)$ & $0.660(8)$ \\
 & & & & & & & & \vspace{-0.32cm} \\
     &       & $(1/2)^+$ & $-$/$\Xi'_c$      & $747(25)$ & $691(29)$ & $-$ & $0.857(22)$ & $-$ \\
     &       & $(3/2)^+$ & $-$/$\Xi_c(2645)$       & $771(25)$ & $715(29)$ & $-$ & $0.886(21)$ & $-$ \\
 & & & & & & & & \vspace{-0.32cm} \\
     &       & $(1/2)^-$ & $-$/$-$                 & $-$ & $-$ & $-$ & $-$ & $-$ \\
 & & & & & & & & \vspace{-0.32cm} \\
     &       & $(1/2)^-$ & $-$/$\Xi_c(2790)$       & $1013(46)$ & $936(48)$ & $-$ & $1.160(45)$ & $-$ \\
     &       & $(3/2)^-$ & $-$/$\Xi_c(2815)$       & $1023(46)$ & $946(48)$ & $-$ & $1.172(45)$ & $-$ \\
 & & & & & & & & \vspace{-0.40cm} \\
\hline
% %%%%%%%%%%
% \Omega
% %%%%%%%%%%
 & & & & & & & & \vspace{-0.40cm} \\
$-2$ & $0$ & $(1/2)^+$ & $\Omega_b$/$\Omega_c$ & $872(25)$ & $807(31)$ & $775(8)$& $1$ & $1$ \\
$ $ & $ $ & $ (3/2)^+$ & $-$/$\Omega_c(2770)$    & $905(25)$ & $839(31)$ & $-$& $\textit{1.030(2)}$ $^2$ & $-$ \\
 & & & & & & & & \vspace{-0.32cm} \\
    &     & $(1/2)^-$ & $-$/$-$                     & $-$ & $-$ & $-$ & $-$ & $-$ \\
 & & & & & & & & \vspace{-0.32cm} \\
    &     & $(1/2)^-$ & $-$/$-$                     & $-$ & $-$ & $-$ & $-$ & $-$ \\
    &     & $(3/2)^-$ & $-$/$-$                     & $-$ & $-$ & $-$ & $-$ & $-$\vspace{-0.40cm} \\
 & & & & & & & & \\
\hline
\end{tabular}
\caption{\label{TAB192}$b$ baryon mass differences $\Delta m(S,I,J^\mathcal{P}) = m(\textrm{baryon}:S,I,J^\mathcal{P}) - m(B)$ in $\textrm{MeV}$ (scale setting via $m_\pi$ and $f_\pi$, $a = 0.079(3) \, \textrm{fm}$ \cite{Baron:2009wt}, and via $m_\pi$ and $m_N$, $a = 0.089(5) \, \textrm{fm}$ \cite{Alexandrou:2011db}) and dimensionless ratios of baryon mass differences differences (cf.\ (\ref{EQN693})); $^1$~experimental results have been taken from \cite{PDG} with exception of $m(\Omega_b)$, which has been taken from \cite{Aaltonen:2009ny}; $^2$~this number does not require any lattice result. Lines associated with quantum numbers, where no corresponding $c$ baryons have experimentally been measured are filled with $-$.}
\end{center}
\end{table}

% **********

\subsection{Discussion of possible systematic errors}

Our lattice results might be associated with certain systematic errors, which we list and briefly discuss in the following.
\begin{itemize}
\item Scale setting: \\ the dominating source of systematic error arises from the ambiguity introduced by the two methods of scale setting \cite{Baron:2009wt,Alexandrou:2011db}, which is around 10\%. Although it seems that the lattice spacing $a = 0.089(5) \, \textrm{fm}$ determined by means of the nucleon mass seems more appropriate, when comparing to experimental results or to publications from other lattice collaborations (cf.\ section~\ref{SEC461} and section~\ref{SEC593}), we strongly recommend to consider the dimensionless ratios $R^{\textrm{lat},b}$, where scale setting errors are essentially eliminated.

\item Extrapolation to the physical $u/d$ quark mass: \\ as explained in section~\ref{SEC486} our results, which cover pion masses in the range $340 \, \textrm{MeV} \ltapprox m_\textrm{PS} \ltapprox 525 \, \textrm{MeV}$, are consistent with a linear dependence in $(m_\textrm{PS})^2$ for all static-light baryon states. Whether there are deviations at significantly lighter $u/d$ quark masses, will be studied using corresponding ETMC gauge field configurations, which will be available soon.

\item Possibly incorrect tuning of the $s$ quark mass: \\ this issue has already been discussed in section~\ref{SEC394}, where we estimate the systematic error for static-light baryons with a single valence $s$ quark to be around $10 \, \textrm{MeV}$ and for those with two valence $s$ quarks to be around $20 \, \textrm{MeV}$. Note that the extrapolation to the physical $b$ quark mass by means of experimental results on $c$ baryons reduces these errors by around $1/3$.

% m(B^*) - m(B) = 46 MeV
% m(B_2^*) - m(B_1) = 19 MeV
% m(\Sigma_b^*) - m(\Sigma_b) = 21 MeV
%
\item Extrapolation to the physical $b$ quark mass: \\ the validity of the interpolation between static lattice results and charm experimental results by the order $1/m_Q$ of HQET has been tested for baryons via $\Sigma_b / \Sigma_b^\ast$ as explained above and for mesons via $B / B^\ast$ \cite{:2010iv}. These two tests indicate validity up to $\approx 5 \%$. Since the hyperfine splitting in the $b$ region is of order $20 \, \textrm{MeV}$ to $50 \, \textrm{MeV}$ (cf.\ e.g.\ $B$/$B^\ast$, $B_1$/$B_2^\ast$, $\Sigma_b$/$\Sigma_b^\ast$ in \cite{PDG}) one expects a corresponding systematic error of $\ltapprox 2.5 \, \textrm{MeV}$.

\item Electromagnetic and isospin breaking effects: \\ experimental results on $\Sigma_b^-$ and $\Sigma_b^+$ indicate that such effects can be of order $5 \, \textrm{MeV}$ to $10 \, \textrm{MeV}$.

\item Neglect of $s$ and $c$ sea quarks: \\ the systematic error arising from our neglect of the $s$ and $c$ quark contribution to the sea is expected to be significantly smaller than current statistical errors. Will will address and quantify this error in a future study making use of recently generated $N_f = 2+1+1$ ETMC gauge field configurations \cite{Baron:2010bv,Baron:2011sf}.

\item Continuum limit: \\ since we use a rather fine lattice spacing and an $\mathcal{O}(a)$ improved lattice formulation (twisted mass lattice QCD at maximal twist), we expect discretization effects to be negligible. This expectation is supported by the computation and comparison of $I = 1$ states ($I_z = \pm 1$ versus $I_z = 0$), which are not degenerate in twisted mass lattice QCD, but differ by $\mathcal{O}(a^2)$. This constitutes a direct check of lattice discretization effects, for which we found no indication (cf.\ Table~\ref{TAB191}). Moreover, we have recently investigated the continuum limit for $b$ mesons using the same gauge field configurations and also did not find any sign of discretization effects \cite{:2010iv}.

\item Multiparticle states: \\ contamination of static-light baryon states by multiparticle states of the same quantum number have been discussed in section~\ref{SEC394}. It seems rather unlikely that they introduce a systematic error, which is significant compared to current statistical errors.
\end{itemize}
%
% estimate the total systematic error:
% - u/d extrapolation and incorrect s quark mass: 20 MeV
% - b interpolation: 2.5 MeV
% - continuum limit: negligible
% - neglect of s/c sea: negligible
% - electromagnetic, isospin: 10 MeV
% - multiparticle: 10 MeV (as for mesons)
% --> sqrt(20^2 + 2.5^2 + 10^2 + 10^2) = 24.6 MeV
%
% systematic error for ratios:
% m \pm dm / (M \pm dM) = m/M \pm dm//M \pm (m dM)/M^2 = 0.0456...
% (m = M = 775 MeV (\Omega_b mass), dm = dM = 25 MeV)
%
In total the sum of these systematic errors should not exceed $25 \, \textrm{MeV}$, which is of the same order of magnitude as for our recent lattice results on $B$ mesons \cite{:2010iv}, where we quoted a maximal systematic error of $20 \, \textrm{MeV}$. An additional uncertainty of 10\% should be assigned, when considering mass differences in $\textrm{MeV}$, i.e.\ $\Delta m^{\textrm{lat},b}(S,I,j^\mathcal{P})$. For the dimensionless ratios $R^{\textrm{lat},b}(S,I,j^\mathcal{P})$ collected in Table~\ref{TAB192} the latter is not present, while the above mentioned $25 \, \textrm{MeV}$ translate to a systematic error of around 5\%.

% **********

\subsection{\label{SEC593}Comparison to experimental results}

In experiments five $b$ baryon states have been measured: $\Lambda_b$, $\Sigma_b$, $\Sigma_b^\ast$, $\Xi_b$ and $\Omega_b$. We compare our lattice results with these experimental results in Table~\ref{TAB192}. As already mentioned in the previous section the lattice spacing depends to some extent on the observables used to introduce physical units. While setting the scale via $m_\pi$ and $f_\pi$ \cite{Baron:2009wt} yields lattice results, which are around 10\% to 20\% larger than their experimental counterparts, using $m_\pi$ and $m_N$ \cite{Alexandrou:2011db} leads to significantly better agreement. To reduce scale setting effects as much as possible, we prefer to compare the dimensionless ratios $R^{\textrm{lat},b}(S,I,j^\mathcal{P})$ and $R^{\textrm{exp},b}(S,j^\mathcal{P},I)$, which have been defined in (\ref{EQN693}). While $R^{\textrm{exp},b}(S,I,j^\mathcal{P})$ denotes the ratio of experimentally measured $b$ baryons, $R^{\textrm{lat},b}(S,I,j^\mathcal{P})$ is the linear $m_c/m_Q$ interpolation between the static-light ratio from Table~\ref{TAB191} and the corresponding ratio of experimentally measured $c$ baryons. As can be seen from Table~\ref{TAB192}, there is reasonable agreement between our lattice and experimental results for the four available ratios.

There are seven more $b$ baryon states we are able to predict, but which have not yet been measured by experiment. Their values in $\textrm{MeV}$ as well as the ratios $R^{\textrm{lat},b}(S,I,j^\mathcal{P})$ are also collected in Table~\ref{TAB192}.

% ********************
% ********************
% ********************
% ********************
% ********************

\newpage

\section{\label{SEC571}Conclusions}

We have computed the spectrum of static-light baryons by means of lattice QCD using $N_f = 2$ flavors of light quarks. We have considered all possible combinations of two light quarks, i.e.\ $\Lambda$, $\Sigma$, $\Xi$ and $\Omega$ baryons, angular momentum/spin of the light degrees of freedom $j \in \{ 0 \, , 1 \}$ and both parity $\mathcal{P} = +$ and $\mathcal{P} = -$. In particular we were able to predict a number of negative parity states, which have at the moment neither been measured experimentally nor previously been computed on the lattice.

We have employed the assumption of a $1/m_Q$ dependence on the heavy quark mass together with experimental results for $c$ baryons to allow us to estimate the spectrum that one would obtain for $b$ quarks of finite physical mass.

The wide variety of computed states (both static-light baryons and $b$ baryons) will be a valuable resource for model builders and might give input for future experiments.

Obvious directions to continue this research include 
(i)~investigating the continuum limit; 
(ii)~performing similar computations at lighter $u/d$ quark masses; 
(iii)~studying the dependence of $\Xi$ and $\Omega$ baryons on the $s$ quark mass; 
(iv)~extending these computations to $N_f = 2+1+1$ flavor ETMC gauge field configurations \cite{Baron:2010bv,Baron:2011sf}; 
(v)~considering non-trivial gluonic excitations allowing to study total angular momentum of the light degrees of freedom $j > 1$; 
(vi)~replacing experimental input for $c$ baryons by corresponding lattice results with heavy quarks of finite mass \cite{Papinutto:2010cb} and/or combining such results with a recently proposed method for lattice $B$ physics \cite{Blossier:2009hg} to compute the spectrum of $b$ baryons in an alternative way.
% (vi)~replacing experimental input for $c$ baryons by corresponding lattice results \cite{Papinutto:2010cb}.

% ********************
% ********************
% ********************
% ********************
% ********************

\appendix

\newpage

\section{\label{APP001}$\Delta m^\textrm{stat}(S,j^\mathcal{P},I) a$ for all four ensembles}

% \begin{table}[htp]
% \begin{center}
\begin{tabular}{|c|c|c|c||c|c|c|c|}
\hline
 & & & & & & & \vspace{-0.40cm} \\
    &     &                         &      & $\Delta m^\textrm{stat} a$, & $\Delta m^\textrm{stat} a$, & $\Delta m^\textrm{stat} a$, & $\Delta m^\textrm{stat} a$ \\
$S$ & $I$ & $j^\mathcal{P}$  & flavor & $\mu_\mathrm{q} = 0.0040$   & $\mu_\mathrm{q}=0.0064$  & $\mu_\mathrm{q}=0.0085$ & $\mu_\mathrm{q}=0.0100$ \\
 & & & & & & & \vspace{-0.40cm} \\
\hline
 & & & & & & & \vspace{-0.40cm} \\
$0$ & $0$ & $0^+$         & $ud-du$      & $0.1889(85)\phantom{0}$ & $0.1845(147)$  & $0.2006(103)$ & $0.2126(96)\phantom{0}$\\
 & & & & & & & \vspace{-0.32cm} \\
    &     & $0^-$         & $ud-du$      & $0.5612(318)$      & $0.4635(898)$             & $0.5893(600)$  & $0.4656(511)$\\

& & & & & & & \vspace{-0.32cm} \\
    &     & $1^-$         & $ud-du$      & $0.3727(175)$      & $0.4425(490)$             & $0.4938(415)$  & $-$ \\
 & & & & & & & \vspace{-0.40cm} \\
\hline
 & & & & & & & \vspace{-0.40cm} \\
$0$ & $1$ & $0^-$         & $ud+du$ & $0.3519(440)$ & $0.3878(635)$           & $0.3336(516)$  & $0.3524(291)$ \\
    &     &              & $uu$/$dd$ & $0.4252(344)$ & $0.3627(429)$ &$0.4621(340)$ & $0.4327(511)$   \\
 & & & & & & & \vspace{-0.32cm} \\
    &     & $1^+$        & $ud+du$      & $0.2629(84)\phantom{0}$ & $0.2891(108)$  & $0.2938(134)$  & $0.2940(146)$\\
    &     &              & $uu$/$dd$ & $0.2697(79)\phantom{0}$ & $0.2696(194)$  &  $0.2777(121)$ & $ 0.2988(128)$\\
 & & & & & & & \vspace{-0.32cm} \\
    &     & $1^-$         & $ud+du$      & $0.4376(162)$      & $ 0.4365(393)$  & $0.5141(472)$ &$0.4616(314)$\\
    &     &               & $uu$/$dd$ & $0.4335(236)$           & $0.4473(495)$            & $ 0.5380(371)$  & $ 0.4856(423)$\\
 & & & & & & & \vspace{-0.40cm} \\
\hline
 & & & & & & & \vspace{-0.40cm} \\
$-1$ & $1/2$ & $0^+$      & $us^-$/$ds^+$ & $0.2419(54)\phantom{0}$ & $0.2346(121)$ & $0.2356(77)\phantom{0}$ &$0.2444(77)\phantom{0}$\\
     &       &            & $us^+$/$ds^-$ & $0.2560(76)\phantom{0}$ & $0.2663(86)\phantom{0}$  & $0.2628(101)$ &$0.2671(105)$\\
 & & & & & & & \vspace{-0.32cm} \\
     &       & $0^-$      & $us^-$/$ds^+$ & $0.4559(247)$      & $0.4065(445)$           & $0.4112(537)$ & $0.4048(298)$\\
     &       &            &  $us^+$/$ds^-$ & $0.4118(320)$           & $0.4139(355)$         & $0.4130(506)$  & $0.4879(114)$\\
 & & & & & & & \vspace{-0.32cm} \\
     &       & $1^+$       & $us^-$/$ds^+$ & $0.3107(53)\phantom{0}$ & $0.3198(79)\phantom{0}$ & $0.3120(91)\phantom{0}$ &$0.3203(104)$\\
     &       &             & $us^+$/$ds^-$ & $0.3131(48)\phantom{0}$ & $0.3119(123)$ & $0.3066(99)\phantom{0}$  &$0.3228(117)$\\
 & & & & & & & \vspace{-0.32cm} \\
     &       & $1^-$       & $us^-$/$ds^+$ & $0.4399(122)$   & $0.4772(242)$ & $0.5113(308)$ & $0.4568(206)$\\
     &       &             & $us^+$/$ds^-$ & $0.4554(177)$    & $0.4666(312)$& $0.5134(349)$ &$0.4923(275)$\\
 & & & & & & & \vspace{-0.40cm} \\
\hline
 & & & & & & & \vspace{-0.40cm} \\
$-2$ & $0$ & $0^-$         & $s^+s^+$/$s^-s^-$ & $0.5195(90)\phantom{0}$  & $0.5070(143)$  & $0.5198(122)$ & $0.4879(114)$\\

     &     &               & $s^+s^-$/$s^-s^+$ & $0.4887(176)$ & $0.4927(364)$ &  $0.4790(455)$& $0.4397(260)$\\
 & & & & & & & \vspace{-0.32cm} \\
     &     & $1^+$         & $s^+s^+$/$s^-s^-$ & $0.3508(34)\phantom{0}$ & $0.3488(86)\phantom{0}$ & $0.3357(80)\phantom{0}$ & $0.3422(94)\phantom{0}$\\
     &     &               & $s^+s^-$/$s^-s^+$ & $0.3513(35)\phantom{0}$ & $0.3488(64)\phantom{0}$ & $0.3349(68)\phantom{0}$ & $0.3451(80)\phantom{0}$\\
 & & & & & & & \vspace{-0.32cm} \\
     &     & $1^-$         &$s^+s^+$/$s^-s^-$ & $0.5150(117)$    & $0.5177(287)$  & $0.5650(235)$ &$0.5281(272)$\\
    &     &             & $s^+s^-$/$s^-s^+$ & $0.5165(75)\phantom{0}$   & $0.5300(219)$       & $0.5460(138)$ &  $0.5279(180)$\vspace{-0.40cm} \\
 & & & & & & & \\
\hline
\end{tabular}
% \caption{$\Delta m^\textrm{stat}(S,j^\mathcal{P},I) a$ for all four ensembles.}
% \end{center}
% \end{table}

% ********************
% ********************
% ********************
% ********************
% ********************

\newpage

\section*{Acknowledgments}

It is a pleasure to thank Vladimir Galkin and Chris Michael for many hours of helpful discussions. We acknowledge further useful discussions with Jaume Carbonell, William Detmold, Karl Jansen, Andreas Kronfeld and Michael M\"uller-Preussker.

The major part of computations has been performed at the PC farm at DESY Zeuthen. We thank DESY as well as its staff for technical advice and help.

This work has been supported in part by the DFG Sonderforschungsbereich TR9 Computer\-gest\"utzte Theoretische Teilchenphysik.

% ********************
% ********************
% ********************
% ********************
% ********************

% ********************


\begin{thebibliography}{99}

\bibitem{:2007rw}
  T.~Aaltonen {\it et al.} [CDF Collaboration],
  ``First observation of heavy baryons $\Sigma_b$ and $\Sigma_b^\ast$,''
  Phys.\ Rev.\ Lett.\ {\bf 99}, 202001 (2007)
  [arXiv:0706.3868 [hep-ex]].
  %%CITATION = PRLTA,99,202001;%%

\bibitem{:2007ub}
  V.~M.~Abazov {\it et al.} [D0 Collaboration],
  ``Direct observation of the strange $b$ baryon $\Xi_b^-$,''
  Phys.\ Rev.\ Lett.\ {\bf 99}, 052001 (2007)
  [arXiv:0706.1690 [hep-ex]].
  %%CITATION = PRLTA,99,052001;%%

\bibitem{:2007un}
  T.~Aaltonen {\it et al.} [CDF Collaboration],
  ``Observation and mass measurement of the baryon $\Xi_b^-$,''
  Phys.\ Rev.\ Lett.\ {\bf 99}, 052002 (2007)
  [arXiv:0707.0589 [hep-ex]].
  %%CITATION = PRLTA,99,052002;%%

\bibitem{Abazov:2008qm}
  V.~M.~Abazov {\it et al.} [D0 Collaboration],
  ``Observation of the doubly strange $b$ baryon $\Omega_b^-$,''
  Phys.\ Rev.\ Lett.\ {\bf 101}, 232002 (2008)
  [arXiv:0808.4142 [hep-ex]].
  %%CITATION = PRLTA,101,232002;%%

\bibitem{Aaltonen:2009ny}
  T.~Aaltonen {\it et al.} [CDF Collaboration],
  ``Observation of the $\Omega_b$-baryon and measurement of the properties of the $\Xi_b$- and $\Omega_b$-baryons,''
  Phys.\ Rev.\ D {\bf 80}, 072003 (2009)
  [arXiv:0905.3123 [hep-ex]].
  %%CITATION = PHRVA,D80,072003;%%

\bibitem{Michael:1998sg}
  C.~Michael and J.~Peisa [UKQCD Collaboration],
  ``Maximal variance reduction for stochastic propagators with applications to the static quark spectrum,''
  Phys.\ Rev.\ D {\bf 58}, 034506 (1998)
  [arXiv:hep-lat/9802015].
  %%CITATION = PHRVA,D58,034506;%%

\bibitem{Detmold:2007wk}
  W.~Detmold, K.~Orginos and M.~J.~Savage,
  ``$BB$ potentials in quenched lattice QCD,''
  Phys.\ Rev.\ D {\bf 76}, 114503 (2007)
  [arXiv:hep-lat/0703009].
  %%CITATION = PHRVA,D76,114503;%%

\bibitem{Burch:2008qx}
  T.~Burch, C.~Hagen, C.~B.~Lang, M.~Limmer and A.~Schafer,
  ``Excitations of single-beauty hadrons,''
  Phys.\ Rev.\ D {\bf 79}, 014504 (2009)
  [arXiv:0809.1103 [hep-lat]].
  %%CITATION = PHRVA,D79,014504;%%

\bibitem{Detmold:2008ww}
  W.~Detmold, C.~J.~Lin and M.~Wingate,
  ``Bottom hadron mass splittings in the static limit from 2+1 flavour lattice QCD,''
  Nucl.\ Phys.\ B {\bf 818}, 17 (2009)
  [arXiv:0812.2583 [hep-lat]].
  %%CITATION = NUPHA,B818,17;%%

\bibitem{Lin:2009rx}
  H.~W.~Lin, S.~D.~Cohen, N.~Mathur and K.~Orginos,
  ``Bottom-hadron mass splittings from static-quark action on $2+1$-flavor lattices,''
  Phys.\ Rev.\ D {\bf 80}, 054027 (2009)
  [arXiv:0905.4120 [hep-lat]].
  %%CITATION = PHRVA,D80,054027;%%

\bibitem{Lin:2010wb}
  H.~W.~Lin, S.~D.~Cohen, L.~Liu, N.~Mathur, K.~Orginos and A.~Walker-Loud,
  ``Heavy-Baryon Spectroscopy from Lattice QCD,''
  Comput.\ Phys.\ Commun.\ {\bf 182}, 24 (2011)
  [arXiv:1002.4710 [hep-lat]].
  %%CITATION = CPHCB,182,24;%%

\bibitem{Neubert:1993mb}
  M.~Neubert,
  ``Heavy quark symmetry,''
  Phys.\ Rept.\ {\bf 245}, 259 (1994)
  [arXiv:hep-ph/9306320].
  %%CITATION = PRPLC,245,259;%%

\bibitem{Mannel:1997ky}
  T.~Mannel,
  ``Heavy-quark effective field theory,''
  Rept.\ Prog.\ Phys.\ {\bf 60}, 1113 (1997).
  %%CITATION = RPPHA,60,1113;%%

\bibitem{Lewis:2008fu}
  R.~Lewis and R.~M.~Woloshyn,
  ``Bottom baryons from a dynamical lattice QCD simulation,''
  Phys.\ Rev.\ D {\bf 79}, 014502 (2009)
  [arXiv:0806.4783 [hep-lat]].
  %%CITATION = PHRVA,D79,014502;%%

\bibitem{Na:2008hz}
  H.~Na and S.~Gottlieb,
  ``Heavy baryon mass spectrum from lattice QCD with $2+1$ dynamical sea quark flavors,''
  PoS {\bf LATTICE2008}, 119 (2008)
  [arXiv:0812.1235 [hep-lat]].
  %%CITATION = POSCI,LATTICE2008,119;%%

\bibitem{Meinel:2009vv}
  S.~Meinel, W.~Detmold, C.~J.~Lin and M.~Wingate,
  ``Bottom hadrons from lattice QCD with domain wall and NRQCD fermions,''
  PoS {\bf LAT2009}, 105 (2009)
  [arXiv:0909.3837 [hep-lat]].
  %%CITATION = POSCI,LAT2009,105;%%

\bibitem{Thacker:1990bm}
  B.~A.~Thacker and G.~P.~Lepage,
  ``Heavy quark bound states in lattice QCD,''
  Phys.\ Rev.\ D {\bf 43}, 196 (1991).
  %%CITATION = PHRVA,D43,196;%%

\bibitem{Lewis:2010xj}
  R.~Lewis,
  ``Bottom and charmed hadron spectroscopy from lattice QCD,''
  arXiv:1010.0889 [hep-lat].
  %%CITATION = ARXIV:1010.0889;%%

\bibitem{Bochicchio:1991cy}
  M.~Bochicchio, G.~Martinelli, C.~R.~Allton, C.~T.~Sachrajda and D.~B.~Carpenter,
  ``Heavy quark spectroscopy on the lattice,''
  Nucl.\ Phys.\ B {\bf 372}, 403 (1992).
  %%CITATION = NUPHA,B372,403;%%

\bibitem{Guazzini:2007bu}
  D.~Guazzini, H.~B.~Meyer and R.~Sommer  [ALPHA Collaboration],
  ``Non-perturbative renormalization of the chromo-magnetic operator in heavy quark effective theory and the $B^\ast$-$B$ mass splitting,''
  JHEP {\bf 0710}, 081 (2007)
  [arXiv:0705.1809 [hep-lat]].
  %%CITATION = JHEPA,0710,081;%%

\bibitem{Blossier:2010jk}
  B.~Blossier, M.~della Morte, N.~Garron and R.~Sommer,
  ``HQET at order $1/m$: I. Non-perturbative parameters in the quenched approximation,''
  JHEP {\bf 1006}, 002 (2010)
  [arXiv:1001.4783 [hep-lat]].
  %%CITATION = JHEPA,1006,002;%%

\bibitem{Blossier:2010vz}
  B.~Blossier, M.~Della Morte, N.~Garron, G.~von Hippel, T.~Mendes, H.~Simma and R.~Sommer [Alpha Collaboration],
  ``HQET at order $1/m$: II. Spectroscopy in the quenched approximation,''
  JHEP {\bf 1005}, 074 (2010)
  [arXiv:1004.2661 [hep-lat]].
  %%CITATION = JHEPA,1005,074;%%

\bibitem{Ebert:2007nw}
  D.~Ebert, R.~N.~Faustov and V.~O.~Galkin,
  ``Masses of excited heavy baryons in the relativistic quark model,''
  Phys.\ Lett.\ B {\bf 659}, 612 (2008)
  [arXiv:0705.2957 [hep-ph]].
  %%CITATION = PHLTA,B659,612;%%

\bibitem{Jansen:2008si}
  K.~Jansen, C.~Michael, A.~Shindler and M.~Wagner [ETM Collaboration],
  ``The static-light meson spectrum from twisted mass lattice QCD,''
  JHEP {\bf 0812}, 058 (2008)
  [arXiv:0810.1843 [hep-lat]].
  %%CITATION = JHEPA,0812,058;%%

\bibitem{:2010iv}
  C.~Michael, A.~Shindler and M.~Wagner [ETM Collaboration],
  ``The continuum limit of the static-light meson spectrum,''
  JHEP {\bf 1008}, 009 (2010)
  [arXiv:1004.4235 [hep-lat]].
  %%CITATION = JHEPA,1008,009;%%

\bibitem{Wagner:2010hj}
  M.~Wagner and C.~Wiese [ETM Collaboration],
  ``The Spectrum of static-light baryons in twisted mass lattice QCD,''
  PoS {\bf LATTICE2010}, 130 (2010)
  [arXiv:1008.0653 [hep-lat]].
  %%CITATION = POSCI,LATTICE2010,130;%%

\bibitem{Weisz:1982zw}
  P.~Weisz,
  ``Continuum limit improved lattice action for pure Yang-Mills theory. 1,''
  Nucl.\ Phys.\ B {\bf 212}, 1 (1983).
  %%CITATION = NUPHA,B212,1;%%

\bibitem{Frezzotti:2000nk}
  R.~Frezzotti, P.~A.~Grassi, S.~Sint and P.~Weisz [Alpha collaboration],
  ``Lattice QCD with a chirally twisted mass term,''
  JHEP {\bf 0108}, 058 (2001)
  [arXiv:hep-lat/0101001].
  %%CITATION = JHEPA,0108,058;%%

\bibitem{Frezzotti:2003ni}
  R.~Frezzotti and G.~C.~Rossi,
  ``Chirally improving Wilson fermions. 1.~$\mathcal{O}(a)$ improvement,''
  JHEP {\bf 0408}, 007 (2004)
  [arXiv:hep-lat/0306014].
  %%CITATION = JHEPA,0408,007;%%

\bibitem{Frezzotti:2004wz}
  R.~Frezzotti and G.~C.~Rossi,
  ``Chirally improving Wilson fermions. 2.~Four-quark operators,''
  JHEP {\bf 0410}, 070 (2004)
  [arXiv:hep-lat/0407002].
  %%CITATION = JHEPA,0410,070;%%

\bibitem{Shindler:2007vp}
  A.~Shindler,
  ``Twisted mass lattice QCD,''
  Phys.\ Rept.\ {\bf 461}, 37 (2008)
  [arXiv:0707.4093 [hep-lat]].
  %%CITATION = PRPLC,461,37;%%

\bibitem{Boucaud:2008xu}
  P.~Boucaud {\it et al.} [ETM Collaboration],
  ``Dynamical twisted mass fermions with light quarks: simulation and analysis details,''
  Comput.\ Phys.\ Commun.\ {\bf 179}, 695 (2008)
  [arXiv:0803.0224 [hep-lat]].
  %%CITATION = CPHCB,179,695;%%

\bibitem{Baron:2009wt}
  R.~Baron {\it et al.} [ETM Collaboration],
  ``Light meson physics from maximally twisted mass lattice QCD,''
  JHEP {\bf 1008}, 097 (2010)
  [arXiv:0911.5061 [hep-lat]].
  %%CITATION = JHEPA,1008,097;%%

\bibitem{Blossier:2007vv}
  B.~Blossier {\it et al.} [ETM Collaboration],
  ``Light quark masses and pseudoscalar decay constants from $N_f = 2$ Lattice QCD with twisted mass fermions,''
  JHEP {\bf 0804}, 020 (2008)
  [arXiv:0709.4574 [hep-lat]].
  %%CITATION = JHEPA,0804,020;%%

\bibitem{Blossier:2009bx}
  B.~Blossier {\it et al.} [ETM Collaboration],
  ``Pseudoscalar decay constants of kaon and $D$-mesons from $N_f = 2$ twisted mass lattice QCD,''
  JHEP {\bf 0907}, 043 (2009)
  [arXiv:0904.0954 [hep-lat]].
  %%CITATION = JHEPA,0907,043;%%

\bibitem{Hasenfratz:2001hp}
  A.~Hasenfratz and F.~Knechtli,
  ``flavour symmetry and the static potential with hypercubic blocking,''
  Phys.\ Rev.\ D {\bf 64}, 034504 (2001)
  [arXiv:hep-lat/0103029].
  %%CITATION = PHRVA,D64,034504;%%

\bibitem{DellaMorte:2003mn}
  M.~Della Morte {\it et al.},
  ``Lattice HQET with exponentially improved statistical precision,''
  Phys. Lett. {\bf B581}, 93, (2004)
  [arXiv:hep-lat/0307021].
  %%CITATION = HEP-LAT/0307021;%%

\bibitem{Della Morte:2005yc}
  M.~Della Morte, A.~Shindler and R.~Sommer,
  ``On lattice actions for static quarks,''
  JHEP {\bf 0508}, 051 (2005)
  [arXiv:hep-lat/0506008].
  %%CITATION = JHEPA,0508,051;%%

\bibitem{Michael:1985ne}
  C.~Michael,
  ``Adjoint sources in lattice gauge theory,''
  Nucl.\ Phys.\ B {\bf 259}, 58 (1985).
  %%CITATION = NUPHA,B259,58;%%

\bibitem{Blossier:2009kd}
  B.~Blossier, M.~Della Morte, G.~von Hippel, T.~Mendes and R.~Sommer,
  ``On the generalized eigenvalue method for energies and matrix elements in lattice field theory,''
  JHEP {\bf 0904}, 094 (2009)
  [arXiv:0902.1265 [hep-lat]].
  %%CITATION = JHEPA,0904,094;%%

\bibitem{Baron:2010th}
  R.~Baron {\it et al.} [ETM Collaboration],
  ``Computing $K$ and $D$ meson masses with $N_f = 2+1+1$ twisted mass lattice QCD,''
  Comput.\ Phys.\ Commun.\ {\bf 182}, 299 (2011)
  [arXiv:1005.2042 [hep-lat]].
  %%CITATION = CPHCB,182,299;%%

\bibitem{Blossier:2009gd}
  B.~Blossier {\it et al.} [ETM Collaboration],
  ``$f_B$ and $f_{B_s}$ with maximally twisted Wilson fermions,''
  PoS {\bf LAT2009}, 151 (2009)
  [arXiv:0911.3757 [hep-lat]].
  %%CITATION = POSCI,LAT2009,151;%%

\bibitem{Alexandrou:2011db}
  C.~Alexandrou {\it et al.} [ETM Collaboration],
  ``Nucleon electromagnetic form factors in twisted mass lattice QCD,''
  arXiv:1102.2208 [hep-lat].
  %%CITATION = ARXIV:1102.2208;%%

\bibitem{McNeile:2000xx}
  C.~McNeile and C.~Michael [UKQCD Collaboration],
  ``Mixing of scalar glueballs and flavour-singlet scalar mesons,''
  Phys.\ Rev.\ D {\bf 63}, 114503 (2001)
  [arXiv:hep-lat/0010019].
  %%CITATION = PHRVA,D63,114503;%%

\bibitem{McNeile:2002az}
  C.~McNeile, C.~Michael and P.~Pennanen [UKQCD Collaboration],
  ``Hybrid meson decay from the lattice,''
  Phys.\ Rev.\ D {\bf 65}, 094505 (2002)
  [arXiv:hep-lat/0201006].
  %%CITATION = PHRVA,D65,094505;%%

\bibitem{McNeile:2004rf}
  C.~McNeile, C.~Michael and G.~Thompson [UKQCD Collaboration],
  ``Hadronic decay of a scalar $B$ meson from the lattice,''
  Phys.\ Rev.\ D {\bf 70}, 054501 (2004)
  [arXiv:hep-lat/0404010].
  %%CITATION = PHRVA,D70,054501;%%

\bibitem{PDG}
  K.\ Nakamura {\it et al.} [Particle Data Group],
  J.\ Phys.\ G {\bf 37}, 075021 (2010).

\bibitem{Baron:2010bv}
  R.~Baron {\it et al.},
  ``Light hadrons from lattice QCD with light $(u,d)$, strange and charm dynamical quarks,''
  JHEP {\bf 1006}, 111 (2010)
  [arXiv:1004.5284 [hep-lat]].
  %%CITATION = JHEPA,1006,111;%%

\bibitem{Baron:2011sf}
  R.~Baron {\it et al.},
  ``Light hadrons from $N_f = 2+1+1$ dynamical twisted mass fermions,''
  PoS {\bf LATTICE2010}, 123 (2010)
  [arXiv:1101.0518 [hep-lat]].
  %%CITATION = POSCI,LATTICE2010,123;%%

\bibitem{Papinutto:2010cb}
  M.~Papinutto, J.~Carbonell, V.~Drach and C.~Alexandrou,
  ``Strange and charmed baryons using $N_f = 2$ twisted mass QCD,''
  PoS {\bf LATTICE2010}, 120 (2010)
  [arXiv:1012.2786 [hep-lat]].
  %%CITATION = POSCI,LATTICE2010,120;%%

\bibitem{Blossier:2009hg}
  B.~Blossier {\it et al.} [ETM Collaboration],
  ``A proposal for $B$-physics on current lattices,''
  JHEP {\bf 1004}, 049 (2010)
  [arXiv:0909.3187 [hep-lat]].
  %%CITATION = JHEPA,1004,049;%%

\end{thebibliography}
\end{document}